\documentclass[
  journal=pasa,
  manuscript=review, 
  year=2025,
  volume=YY,
]{cup-journal}
\usepackage{amsmath}
\usepackage{amssymb,microtype,siunitx,booktabs}
\usepackage{xcolor}
\usepackage{multirow}
\usepackage{hyperref}

\newcommand{\ion}[2]{\mbox{#1\,\textsc{#2}}}

\newcommand{\mghk}[1]{{#1}}

\newcommand{\rn}[1]{{\color{cyan} \bf #1}}

\newcommand{\mghkso}[1]{{\color{gray} \sout{#1} }}
\renewcommand{\mghkso}[1]{}
\newcommand{\arcsec}{^{\prime\prime}}

\newcommand{\rev}[1]{{ #1}}

\title{Evidence for Supermassive Black Hole Binaries}

\author{Martin G. H. Krause}
\affiliation{Centre for Astrophysics Research, Department of Physics, Astronomy and Mathematics, University of Hertfordshire, College Lane, Hatfield, Hertfordshire
AL10 9AB, UK}
\email[Martin Krause]{m.g.h.krause@herts.ac.uk}

\author{Martin A. Bourne}
\affiliation{Centre for Astrophysics Research, Department of Physics, Astronomy and Mathematics, University of Hertfordshire, College Lane, Hatfield, Hertfordshire AL10 9AB, UK}
\alsoaffiliation{Institute of Astronomy, Madingley Road, Cambridge CB3 0HA, UK}
\alsoaffiliation{Kavli Institute for Cosmology, University of Cambridge, Madingley Road, Cambridge CB3 0HA, UK}

\author{Silke Britzen}
\affiliation{Max-Panck-Institut f\"ur Radioastronomie, Auf dem H\"ugel 69, 53121 Bonn, Germany}

\author{Adi Foord}
\affiliation{Kavli Institute of Particle Astrophysics and Cosmology, Stanford University, Stanford, CA 94305, USA}
\alsoaffiliation{Department of Physics, University of Maryland Baltimore County, 1000 Hilltop Cir, Baltimore, MD 21250, USA}

\author{Jenny E. Greene}
\affiliation{Department of Astrophysical Sciences, Princeton University, Princeton, NJ 08544, USA}

\author{Melanie Habouzit}
\affiliation{Max-Planck-Institut f\"ur Astronomie, K\"onigstuhl 17, D-69117 Heidelberg, Germany}
\alsoaffiliation{Zentrum f\"ur Astronomie der Universit\"at Heidelberg, ITA, Albert-Ueberle-Str. 2, D-69120 Heidelberg, Germany}
\alsoaffiliation{University of Geneva, Department of Astronomy, Chemin Pegasi, 51, CH-1290 Versoix, Switzerland}

\author{Maya A. Horton}
\affiliation{Cavendish Astrophysics, University of Cambridge, Cavendish Laboratory, JJ Thomson Avenue Cambridge CB3 0HE, UK}

\makeatletter
\global\cup@affil@cnt=11
\makeatother
\edef\@currentaffil{\number\cup@affil@cnt}

\author{Lucio Mayer}
\affiliation[8]{Department of Astrophysics, University of Zurich, Winterthurerstrasse 190, CH-8057 Z\"urich, Switzerland}

\author{Hannah Middleton}
\affiliation{School of Physics and Astronomy \& Institute for Gravitational Wave Astronomy, University of Birmingham, Birmingham, B15 2TT, UK} 

\author{Rebecca Nealon}
\affiliation{Centre for Exoplanets and Habitability, University of Warwick, Coventry CV4 7AL, UK}
\alsoaffiliation{Department of Physics, University of Warwick, Coventry CV4 7AL, UK}

\author{Julia M. Sisk-Reyn{\'{e}}s}
\affiliation{Center for Astrophysics $|$ Harvard \& Smithsonian, Cambridge, MA 02138, USA}
\alsoaffiliation{Institute of Astronomy, Madingley Road, Cambridge CB3 0HA, UK}

\author{Christopher S. Reynolds}
\affiliation{Department of Astronomy, University of Maryland, College Park, MD 20742-2421, USA
Joint Space-Science Institute (JSI), College Park, MD 20742-2421, USA}
\alsoaffiliation{Joint Space-Science Institute (JSI), College Park, MD 20742-2421, USA}
\alsoaffiliation{Institute of Astronomy, Madingley Road, Cambridge CB3 0HA, UK}

\author{Debora Sijacki}
\affiliation{Institute of Astronomy, Madingley Road, Cambridge CB3 0HA, UK}
\alsoaffiliation{Kavli Institute for Cosmology, University of Cambridge, Madingley Road, Cambridge CB3 0HA, UK}

\doi{10.1017/pasa.2020.32}

\received {dd Mmm YYYY}
\revised  {dd Mmm YYYY}
\accepted {dd Mmm YYYY}
\published{22 September 202X}

\begin{document}
\enlargethispage{1.5\baselineskip}
\vspace{-1.5em}
\begin{abstract}
We review the state of the evidence for the existence and observational appearance of supermassive black hole binaries. Such objects are expected from standard hierarchical galaxy evolution to form 
\rev{after two galaxies, each containing a supermassive black hole, have merged, in the centre of the merger remnant.}
A complex interaction is predicted to take place with stars and gas in the host galaxy, leading to observable signatures in weakly as well as actively accreting phases. Direct observational evidence is available and shows examples of dual active galactic nuclei from kpc scales down to parsec scales. Signatures of possibly closer supermassive black hole binaries may be seen in jetted black holes. The interaction with stars and gas in a galaxy significantly affects the hardening of the binary and hence contributes to uncertainties of the expected gravitational wave signal. 
\rev{The Laser Interferometer Space Antenna (LISA) should in the future detect actual mergers. Before the launch of LISA, pulsar timing arrays may have the best chance to detect a gravitational wave signal from supermassive black hole binaries.}
The first signs of the combined background of \rev{inspiralling objects} might have been seen already.
\end{abstract}

\section{Introduction} \label{sec:intro}

The concept of the black hole goes back to
Karl Schwarzschild's point mass solution of 
Albert Einstein's 
field equations of general relativity
\citep{1916AnP...354..769E,1916SPAW.......189S}.
The characteristic of a black hole is a horizon,
a closed surface, from the inside of which
escape would require superluminal motion and is hence
not possible. As a consequence of
general relativity, we expect such horizons 
to form for extremely compact objects, i.e.,
when the size of an object with mass $M$ is 
\rev{smaller than}
its Schwarzschild radius,
$r_\mathrm{s}=2GM/c^2 \approx 3\,$km$\,(M/M_\odot)$, 
where $G$ is the gravitational constant and 
$c$ the speed of light.
Rotating solutions to Einstein's
field equations have been found by 
\citet{1963PhRvL..11..237K}. They also possess
a horizon, if the spin angular momentum is 
below $J_\mathrm{max}=GM^2/c$. Often, a spin parameter $a=J/J_\mathrm{max}$ is used instead, where $a=1$ denotes a maximally rotating black hole. 

That such extremely compact configurations
could indeed exist in nature was first hinted
at by the huge energy requirements
\citep{1956ApJ...124..416B} to power
extragalactic radio sources and jets from
the nuclei of galaxies 
(\citealt{1964ApJ...140..796S}, also compare \citealt{2019ARA&A..57..467B}).
This was corroborated when the high distances and therefore luminosities
of active galactic
nuclei were established \citep{1968ApJ...151..393S}.
With optical luminosities in excess of a hundred times the luminosity of the Milky Way and energy
requirements for radio sources that equated the conversion of sometimes more than a million times the mass of the Sun
to energy, a single compact object of a mass 
comparable to the Sun was clearly insufficient as power source, and consequently, supermassive ($>10^6M_\odot$) black holes (SMBH) were postulated to explain the observations
\citep{1964ApJ...140..796S,1969Natur.223..690L}.

Supermassive black holes influence their environment
via their gravitational attraction. Their effect
on the kinematics of surrounding stars has been
measured in the nuclei of many galaxies
\citep{1995ARA&A..33..581K,1998AJ....115.2285M}, most convincingly in the centre of our 
own galaxy \citep{2008ApJ...689.1044G,2010RvMP...82.3121G}.
The Nobel prize to Andrea Ghez and Reinhard Genzel in 2020 marked the extensive work that has been done to constrain this object of four million
solar masses \citep{2021arXiv210213000G}. 

The most direct evidence for black holes comes
from gravitational waves. Gravitational waves are emitted when two compact objects orbit each other and, particularly, when they finally merge. The detection of gravitational waves thus implies the existence of binaries of compact objects. 
Gravitational waves were directly detected for the first time in 2015, from the merger of a stellar mass black hole
binary \citep{2016PhRvL.116f1102A}, for which 
a 2017 Nobel prize was awarded \citep{2017Natur.550...19C}. Many events have been
detected since, with the most massive black hole
discovered in this way having a mass of 
$150 M_\odot$ \citep{2020PhRvL.125j1102A}.
The physical setup of Advanced LIGO detectors used in these observations is a Michelson interferometer with 4~km arm length. This limits the accessible
frequency range and thus only stellar mass mergers can be observed. 
Direct detections of SMBHs will
therefore have to wait for the space-borne
LISA mission \citep[Laser Interferometer Space Antenna,][]{2023LRR....26....2A} with
arm lengths of $\sim 2.5$\,million km.
In the meantime, pulsar timing arrays accurately monitor the 
\rev{arrival times of pulses from}
networks of nearby pulsars. Probing 
\rev{nanohertz frequencies}
they have the best chance to detect signals from the most massive supermassive black hole binaries (Sect.~\ref{sec:GW}).

There are good reasons to believe
that such supermassive binaries exist \citep[e.g.][]{1980Natur.287..307B}. SMBHs are nowadays thought to be ubiquitous in galaxies. Their evolution is studied as an essential part of galaxy evolution, though still with considerable uncertainties (Sect.~\ref{sec:smbh-growth}). When galaxies merge, which is cosmologically a frequent event, we expect their SMBHs to migrate towards the new centre and for many cases to merge well within the Hubble time (Sect.~\ref{sec:final-pc}). This expectation is supported by a growing amount of direct 
observational evidence, obtained with similar techniques than the ones that have been used to establish the case for supermassive black holes in the first place (compare above). Activity of both black holes in such a system might be expected relatively frequently, as the merger event that brought in the secondary black hole might also bring in fresh gas. Dual active galactic nuclei with separations of the order of kiloparsec have been directly observed in X-rays, and highest resolution radio imaging has found a first good case of radio cores with parsec-scale separation (Sect.~\ref{sec:dual-AGN}). 
Emission line signatures may reveal periodic kinematics in Doppler shifts in the future (Sect.~\ref{sec:lines}).
\rev{
The direction of the black hole spin is the likely direction for jet ejection and likely parsec-scale SMBH binaries 
have been found to show dynamically changing jet features consistent with orbital motion (Sects.~\ref{sec:VLBI}). 
For such a close binary configuration, the relativistic geodetic precession effect would precess both spins on a much longer timescale. 
Corresponding precession features may have been seen on the $\approx 100$~kpc scale (Sect.~\ref{sec:prec-jets}). 
Cosmological simulations predict the magnitude of the black hole spins to be generally high from gas accretion, and to decrease after binary SMBH mergers. This can now be studied via X-ray spectroscopy for a number of accreting objects (Sect.~\ref{sec:spins}). There are hints
for spin-down at the higher masses, as expected from binary SMBH mergers.
When black holes become gravitationally bound in a binary, the accretion flow becomes much more complex than single discs around each object (Sect.~\ref{sec:discs-breaking-and-spin}). This also influences the spin evolution and observational signatures of
the double active galactic nucleus. 
The spin will also affect the signal shape of the gravitational waves in the eventual merger
(Sect.~\ref{sec:GW}). In the following sections, we review the state of these
fields, starting with the area of perhaps greatest uncertainty.
}

\section{The Growth of Supermassive Black Holes over Cosmic Time  } \label{sec:smbh-growth}
Supermassive black holes  are found in the centre of many, if not all, galaxies. In the local Universe, the diverse population of SMBHs encompasses objects with $\sim 10^{5}\, \rm M_{\odot}$ to SMBHs weighing up to $10^{9}-10^{10}\, \rm M_{\odot}$ in the largest galaxies \citep{2020ARA&A..58..257G}. At higher redshifts, \rev{still} only accreting SMBHs (referred to as active galactic nuclei, AGN, 
and as quasars in their most powerful state) can be detected with current electromagnetic telescopes, 
\rev{though, progress in instrumentation is about to change this \citep{2025arXiv250317478N}. }
The discovery of quasars powered by extremely massive SMBHs with $10^{8-10}\, \rm M_{\odot}$ only 700 Myr after the Big Bang \citep[e.g.,][]{2021ApJ...907L...1W} implies that SMBHs must originate from black hole seeds formed in the early Universe that have grown in mass efficiently over time by accretion of gas and stars, and mergers with other SMBHs \rev{\citep{2023ARAA..61..373F}}. Yet, the exact mechanism(s) of SMBH formation and the relative contributions of SMBH growth channels remain outstanding and unanswered questions in modern astrophysics.

\subsection{Supermassive black hole seeding}
Most of today's leading theoretical models form black hole seeds from the collapse of different massive stellar objects in metal-free or metal-poor halos \citep[][for recent reviews]{2020ARA&A..58...27I,2021NatRP...3..732V}. Low-metallicity environments favour the formation of more massive seeds. The reason is twofold: the mass of the stellar object is set by the Jeans mass, the minimum mass a gas cloud must possess
to overcome its gas pressure and collapse due to its own gravity, which increases with temperature. 
In the absence of efficient coolants (molecular hydrogen and metals), the Jeans mass of a collapsing gas cloud, 
which sets the mass of the stellar object that will form, remains high. Secondly, mass loss from that stellar object through winds is proportional to metallicity, and thus
will remain small in metal-poor halos. A SMBH seed of similar mass will form.  
Based on the above, seeds could form from the collapse of the first-generation stars (Pop-III stars), from the collapse of 
\rev{very massive stars}
(VMS) formed by runaway mergers between stars or by hierarchical mergers of low-mass black holes in dense stellar clusters. Under rare conditions, Super Massive Stars (SMS) could form in atomic cooling halos and collapse into massive seeds.
Rare major mergers of high-redshift and very massive galaxies (even in solar-metallicity environments) could both enhance gas inflow to the nuclear region and prevent cooling sufficiently to induce the formation of a SMS or a SMBH directly \citep{2010Natur.466.1082M,2023arXiv230402066M}. Finally, SMBHs could form well before the first stars and galaxies, potentially as early as inflation, if they originate from the collapse of high-contrast density perturbations in the primordial Universe \citep[e.g.][]{2020ARNPS..70..355C}. These SMBHs are referred to as primordial SMBHs.
The channels mentioned above predict various ranges of initial masses (from tens of $\rm M_{\odot}$ to almost $10^{6}\, \rm M_{\odot}$) and different abundances of black hole seeds.

\subsection{Supermassive black hole growth}
Not all seeds will blossom to become galaxies' central engines, and not all SMBH formation models produce enough seeds to explain the presence of SMBHs in all galaxies.
Light seeds with $\leqslant 10^{4-5}\, \rm M_{\odot}$ will struggle to achieve an efficient growth: they can form off the central gas reservoir of their host galaxies, \mghk{and would then }have a hard time sinking to the galaxy centre after galaxy mergers. \mghk{This} will limit both their accretion rates and their growth by SMBH mergers. Light seeds born in stellar clusters could alleviate these hurdles. Heavy seeds have the advantage to form massive but may be too rare to explain the bulk of the SMBH population \citep{2016MNRAS.463..529H}. The growth of SMBHs is a multi-faceted and complex problem \mghk{due} to their interplay with their host galaxies, and their galactic and large-scale environments \citep[replenishment of gas from large-scale filaments, number of galaxy mergers, formation timescale of SMBH binaries set by galaxy properties, see Sec.~\ref{sec:final-pc} and][for a review]{2023LRR....26....2A}.

Over the last decade, the community has led a major effort to numerically address the formation and evolution of galaxies in a cosmological context with large-scale cosmological hydrodynamical simulations. \rev{The computational domain evolved in such simulations is typically a cube} of $100\, \rm comoving \, Mpc$ side length such as e.g., Illustris \citep{2015MNRAS.452..575S}, Horizon-AGN \citep{2016MNRAS.463.3948D}, or the recent Astrid \citep{2022MNRAS.513..670N}. They allow us to capture a large number of highly non-linear physical processes, diverse environments from dense regions to cosmic voids, and to track the cosmic evolution of thousands of galaxies with
stellar masses in the range $10^{9}$ to $10^{12}\, \rm M_{\odot}$.
The resolution of these large-scale simulations is not sufficient to resolve processes across the entire dynamical range needed, from SMBH accretion discs to large-scale filaments. Processes related to SMBH formation, growth, and feedback, as well as any other baryonic processes taking place at small scales below the galactic scale, are modelled with sub-grid physics. Simulations all use different sub-grid models, e.g. different initial location and mass for SMBHs \citep[the physical formation mechanisms described above are not modeled in these simulations, but see][]{2017MNRAS.468.3935H,2017MNRAS.470.1121T}, 
different models to compute the accretion onto SMBHs \citep[e.g., variations of the Bondi models or torque-limited accretion]{2015MNRAS.454.1038R,2015ApJ...800..127A}, different efficiencies and models to release energy from AGN, some models assuming that AGN feedback channels explicitly depend on SMBH mass \citep{2017MNRAS.465.3291W}, some others assuming a uniform feedback \citep{2015MNRAS.446..521S}. Numerical difficulties persist to model accurately SMBH dynamics in these simulations of $1-2 \, \rm kpc$ spatial resolution. Resolution of about $10\, \rm pc$ would be needed to start resolving the dynamical friction from dark matter, gas, and stars, i.e. only the first phase in the long path of SMBH coalescence \citep{2015MNRAS.447.2123C,2019MNRAS.486..101P}. \rev{Such a fine resolution is only feasible in simulations that consider only individual galaxies or individual pairs of interacting galaxies (compare Sect.~\ref{sec:final-pc}).}
Nonetheless, these fascinating multi-scale simulations have been able to reproduce a number of observed galaxy properties (e.g., realistic galaxy and halo stellar mass contents, star formation rates,
galaxy morphologies). However, detailed analyses have shown that these simulations all produce different populations of SMBHs and AGN across cosmic times and highlighted differences with current observational constraints \citep[e.g., with the AGN luminosity function beyond $z=0$, AGN fraction in low-mass galaxies, $M_{\rm MBH}-\sigma$ relation,][]{2021MNRAS.503.1940H,2022MNRAS.509.3015H,2020ApJ...895..102L,2021MNRAS.508.4816S,2022MNRAS.514.4912H}.
In particular, there is no consensus among the simulations on the shape of the $M_{\rm MBH}-M_{\star}$ relation, its scatter in $M_{\rm MBH}$ at fixed $M_{\star}$, and its evolution with redshift. The latter aspect has important implications: half of the large-scale simulations of the field predict that SMBHs are on average more massive with respect to their host galaxies at high redshift (e.g., $z\geqslant 4$) than they are in the local Universe, the other half predict under-massive SMBHs at high redshift. With the advent of the James Webb Space Telescope (JWST), the build-up of the $M_{\rm MBH}-M_{\star}$ relation can for the first time be constrained at $z\geqslant 6$ with the observations of faint quasars \rev{($L_{\rm bol}=10^{44}-10^{46}\, \rm erg/s$)} and their host galaxies\rev{ \citep{2022MNRAS.511.3751H,2024AA...691A.145M,2025ApSS.370...85H}}. JWST enables both the characterization of the stellar component of high-redshift quasar hosts \citep[][for the first detection]{2022arXiv221114329D} and more precise SMBH mass measurements. Such novel constraints will help improve our understanding and modeling of SMBH seeding, SMBH accretion \citep[e.g.,][]{2017MNRAS.467.3475N}, SMBH mergers, and the role of supernovae \citep[e.g.,][]{2017MNRAS.468.3935H} and AGN feedback \citep[e.g.,][]{2022MNRAS.516.2112K} in cosmological simulations. This is crucial as those are all degenerate aspects of SMBH growth. 

While comparing the large-scale cosmological simulations of the field, large discrepancies emerge on the number density and fraction of dual AGN that they predict \citep{2025MNRAS.536.3016P}; reassuring that new observations are needed to constrain the AGN population in simulations. 
As described below dual AGN can be the precursors of SMBH binaries and the future events of the LISA gravitational wave antenna. LISA will provide a unique view of the merger history of black holes in the range $10^{4}-10^{7}\, \rm M_{\odot}$ at any redshift.

\section{Black-hole migration within the remnants of galaxy mergers and the last-parsec problem} \label{sec:final-pc}

When galaxies with embedded SMBHs merge, the first phase of the orbital evolution of the black hole pair is
dictated by the large-scale dynamics of the galaxy merger \citep{mayer07,capelo15}. Initially the  orbital evolution is 
governed by dynamical friction  between the
two dark halos. \mghk{This means that the gravitational interaction between the SMBHs and the dark matter leads to an exchange of energy and angular momentum.} As the galaxies merge, the two black holes eventually end up in the remnant at some separation,
forming a loose SMBH pair. The separation can vary between 100 pc and more than a kpc
depending on the mass ratio between the galaxies, being larger for minor
mergers \citep{callegari09}.
In the latter the secondary galaxy, depending on its central density relative to that of the background galaxy, can be tidally disrupted before reaching the centre of the
primary. If the latter happens, the secondary black hole can be left  "naked" at significant distance from the centre, which results in a long dynamical friction timescale  \citep{callegari09,2018ApJ...857L..22T}.
The dynamics of the SMBH pair embedded in the merger remnant
is  then highly dependent on the structural properties of the host galaxy, in  particular on the nature of the gaseous and stellar 
background providing the source of the drag onto the binary. The
modelling uncertainties generally associated with baryonic processes
in galaxy formation inevitably result in uncertainties in the effects of the gaseous background, while if the decay is governed by the
interaction with stars the available models are much more robust.
As a result, the timescale to form a bound binary, which forms at a separation of order 
1-10 pc, depending on the mass of the black holes, can  vary from less than a few tens of million years
to several billion years.  In the most extreme cases, the secondary black hole will be left wandering in the galaxy merger remnant \citep{Bellovary_et_al_2010,2018ApJ...857L..22T}.
\mghk{In the following, we}
address the current knowledge and
main open problems in understanding the entire evolution of SMBH pairs,
from binary formation to coalescence.

\subsection{Binary hardening on the last parsecs in gas-poor galaxy centres}
The final parsec problem was first described by \cite{1980Natur.287..307B} who identified the fundamental timescales involved in SMBH mergers.  It was subsequently seen in numerical N-Body simulations of massive black holes in gas-free spherically symmetric nuclei \citep{2001ApJ...563...34M}.
The original models were  constructed to mimic the mass distribution in the nuclei
of early-type galaxies, the systems in which  the largest black holes had been previously discovered.  In this regime
the two SMBHs first evolve as a result of dynamical friction against the stellar background, until they
reach a separation at which their total mass supersedes the mass of the stars within their orbit. At this point,
corresponding to a separation of  1-10 pc for black holes with masses
exceeding a million solar masses, dynamical friction becomes inefficient. Instead, 3-body encounters
between the loose SMBH binary and individual stars can drain orbital angular momentum and orbital energy, leading to 
\rev{a decrease of the orbital separation so that the binary becomes more tighly bound, i.e., a}
hardening of the binary. However, in spherical systems only a small region of the galactic nucleus comprises stars that can undergo close encounters with the binary. 
\mghk{An important concept in this context is the {\em loss cone}, a term which describes the positions and velocities stars must have to be ejected by interaction with the supermassive binary. The}
 reservoir of interacting stars steadily decreases as they are ejected after encounters, leading rapidly to an empty loss cone, hence
to the stalling of the orbital decay. It is the
 latter occurrence that was originally referred to as the "last parsec problem" \citep{2001ApJ...563...34M}

Since these early studies, a number of ever more sophisticated N-Body simulations with increasingly high resolution have shown that, once the host potential is axisymmetric or triaxial\mghk{, rather than spherically symmetric},
the loss-cone remains filled owing to plentiful elongated plunging  stellar orbits, such as tube/box orbits \citep{2006ApJ...642L..21B,2013ApJ...773..100K}, thus allowing the binary to harden further. The loss cone refilling is a purely collisionless phenomenon, and indeed
N-body simulations with collisionless gravity solvers, such as tree codes, 
have nearly achieved convergence of  loss cone refilling with increasing resolution \citep{2017MNRAS.464.2301G}. Finally, zoom-in cosmological simulations \mghk{using particles to represent the different kinds of matter in the Universe}, augmented with particle splitting to increase the mass and spatial resolution  further in subsequent steps, have confirmed
that the loss-cone is efficiently refilled in  massive galaxies \citep{2016ApJ...828...73K}. They have further pointed out that the hardening timescale below parsec separations,  down to the milliparsec separation stage at which gravitational wave emission takes over, can be as short as a few tens of Myr for black holes with masses above $10^8$ solar masses  in  galactic nuclei with high central densities matching  those of galaxies at $z > 3$. 
\rev{T}he hardening timescale is in the range $10^8-10^9$ yr in the low density cores of present-day spiral galaxies, becoming strongly dependent on the orbital configuration of the galaxy merger \citep{2018ApJ...868...97K}. While more studies are required to best pin-down the range of hardening timescales for a realistic distribution of galaxy structural parameters and orbital configurations,  it is fair to say that, generically, there is no 
last parsec problem when binary hardening is driven by stars.
This is consistent with the recent determination of the gravitational
wave background by \mghk{pulsar timing arrays (discussed in more detail in Sect.~\ref{sec:GW}, below)}, which is thought to be contributed by very large merging SMBHs hosted in early-type 
galaxies in which hardening is driven by stars.

\subsection{Binary hardening in gas-rich galaxy centres}
\begin{figure}[ht!]
\includegraphics[width=\textwidth]{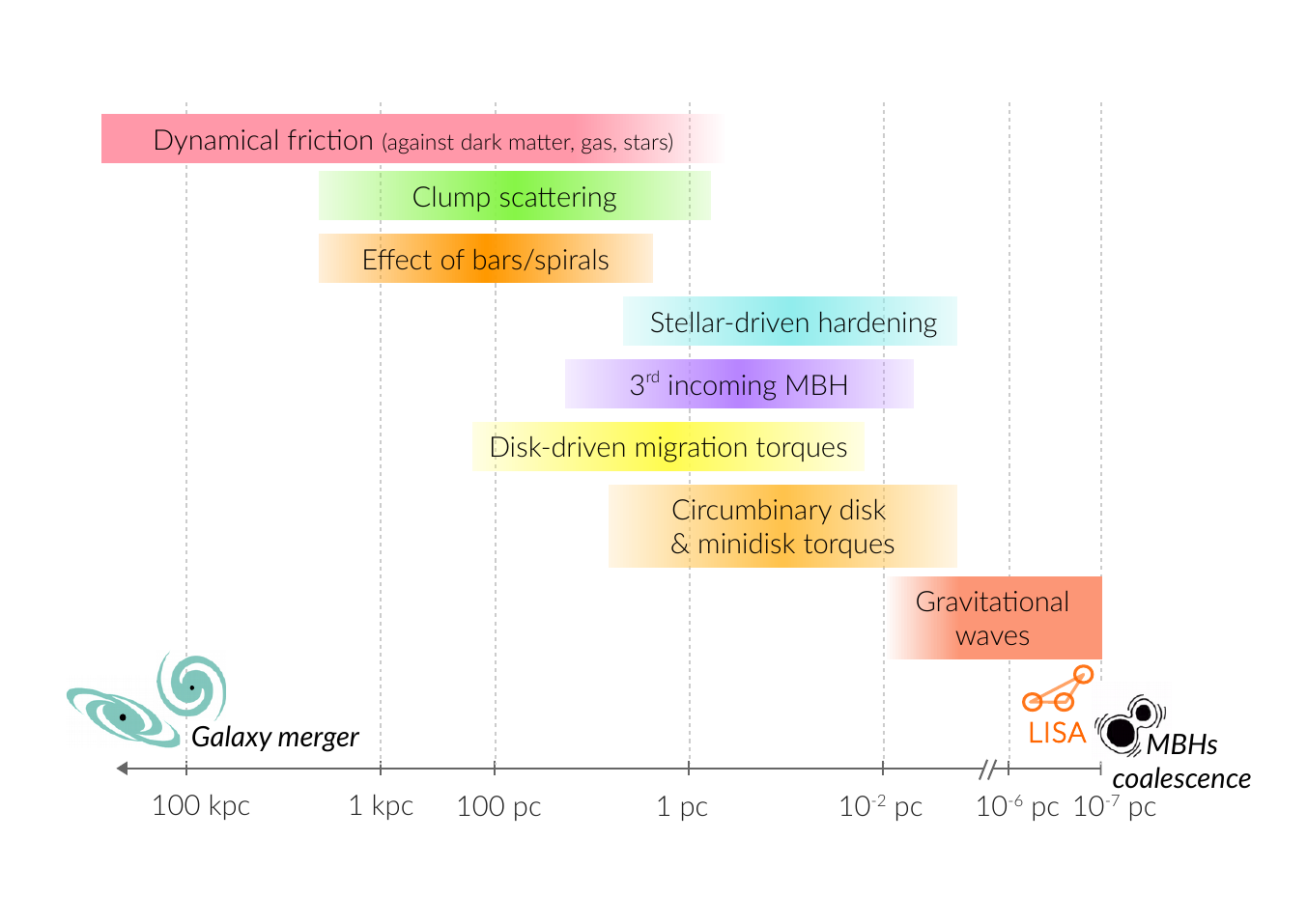}
\caption{A summary of the various astrophysical mechanisms that contribute to drive the orbital evolution of massive black hole pairs 
from large to small scales, adapted from \citet{2023LRR....26....2A} under Creative Commons license, 
\url{https://creativecommons.org/licenses/by/4.0/}.}
\label{fig:decayregimes}
\end{figure}

In gas-rich nuclei the phenomenology of orbital decay is much more complex. While initially it was believed that both SMBH binary formation by dynamical friction and hardening are more efficient in gaseous 
backgrounds rather than in 
stellar-dominated systems \citep{2005ApJ...630..152E, mayer07, 2013CQGra..30x4008M} , numerical
hydrodynamical simulations carried out at various scales have shown that this simple conclusion
applies only to smooth gaseous galactic nuclei, such as axisymmetric nuclear discs or
spherical clouds. Realistic gas-rich galactic nuclei have significant asymmetries, substructure
and clumpiness that can hamper orbital decay at various stages
\citep{2023LRR....26....2A}.
In the following we summarise the main interaction mechanisms between the SMBH binary
and the ambient medium, and how they affect its orbital evolution, eventually suppressing decay, starting with the scale between parsecs and a kiloparsec, and  continuing with evolution at sub-pc scales (compare Fig. ~\ref{fig:decayregimes}).
It is important to underline that the case of a SMBH pair interacting with a predominantly gaseous background is most relevant to LISA sources, since the SMBH mergers with the loudest signals in the LISA band are in a mass range, $10^5-10^7$ solar masses and for such black holes the typical hosts are massive gas-rich spirals or (gas-rich) dwarf galaxies.

Once the two black holes have formed a binary, they find themselves embedded in a dense circumnuclear disc of gas and stars, of order
a hundred pc in size, whose properties
can be inferred from observations of nearby galaxy merger remnants 
\citep{Medling_et_al_2014}.
The gas in the circumnuclear disc is expected to cool radiatively to temperatures 
below $10^4$ K due to fine-structure line and molecular cooling. \mghk{The radiated energy comes from out of the thermal energy reservoir. Slightly denser regions lose their pressure faster, cooling further. The disc is compressed in such regions, thus   } 
becoming \mghk{overall} inhomogeneous and clumpy. The dense clumpy phase corresponds to 
star forming giant molecular clouds (GMCs), weighing a million solar masses, consistent with the high specific star formation rates measured in such systems \citep{Medling_et_al_2014}.
In this regime, scattering between the secondary lighter SMBH and the GMCs becomes dynamically important for SMBHs below $10^7$ solar masses, namely those within the LISA band.  As a result of scattering with GMC-sized clumps, the secondary SMBH can be ejected from the disc midplane,  or undergo episodic outward migration
\citep{2013ApJ...777L..14F,2015MNRAS.449..494R, 2017ApJ...838...13S}.
When it is ejected to several tens of parsecs dynamical friction becomes inefficient as the SMBH finds itself in a low density background (even when accounting for the stellar density of an embedding stellar bulge). As a result the SMBH binary formation timescale becomes long, of order a few 100~Myr as opposed to less than 10 Myr in a smooth disc 
\citep{ 2015MNRAS.449..494R}.
In some cases, though, orbital decay can be accelerated relative to the
smooth background case, because the net torque can still be negative by
chance superposition of multiple torques components generated by 
multiple clumps and spiral density waves  \citep{ 2013ApJ...777L..14F}.
Stochasticity in the outcome of orbital decay is  the most
general trait emerging from numerical simulations of SMBH pairs 
embedded in multi-phase circumnuclear discs.
\rev{By stochasticity we mean that, under small changes of the orbital parameters of the black hole pair
or of the physical parameters defining the properties of the environment (such as density, temperature, or
other properties defining the background gravitational potential), the orbital evolution
of the pair can change drastically. In other words, the trajectory of the black hole pair performs a random
walk in parameter space, so that whether or not the separation of the pair increases or decreases, and
what pace it does that, cannot be predicted.}

Simulations
of massive clumpy galactic discs,  which are common in star forming galaxies at high redshift ($z = 1-3$), have shown that the  same
phenomenology can occur at  larger scales, a few kpc, and for black holes with masses
up to a few times $10^8M_\odot$ . In the latter case the mass of the gaseous clumps is larger
\citep[even exceeding $10^8 M_\odot$,][]{2017MNRAS.464.2952T}, becoming thus
comparable to that of even the most massive SMBHs.  Since the timescales are longer in
a galactic disc compared to a circumnuclear disc, indefinite stalling
is a more likely outcome in these systems. Stochasticity, however, is the prevailing trait also in this case.

If a galactic or circumnuclear disc is not clumpy, the orbital decay
of an SMBH pair is still subject to a complex interaction with the
non-axisymmetric background. In particular, high-resolution  cosmological simulations of  star forming gas-rich disc galaxies have shown that, when the secondary SMBH is delivered at scale of order 0.5-1 kpc, a common occurrence of minor mergers, its further sinking can be hampered as well as promoted by large-scale structure in the merger remnant, such as spiral density waves and bars.
The gravitational torque provided by the latter is found to dominate over the 
local effect of dynamical friction, resulting in a markedly stochastic  orbital evolution \citep{2020MNRAS.498.3601B,2022MNRAS.512.3365B}.

In both circumnuclear discs and galactic discs
the dynamical effect of clumps, spiral structure and bars is mediated by supernovae (SN)  and AGN feedback, which, by heavily impacting both the local and the global density and temperature structure of the ambient medium via heating and wind/outflow driving, affects the magnitude and direction of the torques \citep{2018MNRAS.480..439D}.  In particular, under the simple assumption that the  coupling of AGN feedback with the gas is mainly thermal, the dynamical friction wake can be washed away, leading to the stalling of the SMBH, a phenomenon that has been dubbed "wake evacuation effect" 
\citep{2017ApJ...838...13S, park17}.
Overall, among simulations of SMBHs embedded in clumpy discs, those in which both SN and AGN feedback are included have a higher occurrence of SMBH stalling.

If the SMBHs manage to reduce their separation below parsec scales in a gas-dominated
galactic nucleus, the expectation is that, eventually, a cavity will form around the binary as the surrounding medium absorbs the excess orbital angular momentum and is torqued away. A circumbinary disc forms, and, in absence of self-gravity, the extraction of angular momentum which triggers the decay of the binary can be understood as resulting from a resonant interaction between the binary and the disc \citep{artymowicz1994}. 
Viscous and pressure  forces act against cavity formation, and
can partially refill the cavity, hence this stage is, like the others, dependent on the details of gas physics. Most of the work carried out in this regime  assumes  a locally isothermal or adiabatic medium, neglects self-gravity, accretion and/or feedback from the SMBHs \citep[see the introductory section of][]{Duffell:2024}. Most often a fixed binary orbit is assumed, and the mass ratio considered for the two SMBHs is in the range $0.1-1$. 
These limitations in the physical modelling and parameter space call for caution in the generalisation of the results.  A large variety of numerical hydro codes,
from smooth particle hydrodanamics over static grid-based, to quasi-Lagrangian codes such as moving mesh
and the meshless finite mass method (MFM), has been or is being employed
in this regime. A major code comparison, initiated at KITP during the
Spring 2022 (BINARY22 program) has been recently completed \citep{Duffell:2024}.

From the existing literature, some clear trends have emerged.  As for the 
regimes at larger scales, torques \rev{acting on the black hole binary} can be both negative and positive.
First, in cold discs with low aspect ratio ($H/r < 0.05$, \rev{where $H$ is the disk scale height and $r$
the radial scale length of the disk}) migration \rev{of the black hole binary} is generally inward  \citep{tiede2020,2022ApJ...929L..13F}, while it can be outward otherwise \citep{MunozEtAl2019,DuffellEtAl2019}. The outward migration occurs due to a positive torque predominantly generated by the mini-discs of gas surrounding the two SMBHs inside the cavity. This is also very sensitive to viscosity as the latter controls the amount of material flowing through the cavity, as well as the rate at which it flows  \citep{heathnixon2020,2022ApJ...929L..13F}.

When self-gravity is included, i.e., where the gas mass becomes non-negligible, the regime of the
torques changes again \citep{Cuadra2009}. The torque is found to be negative because it is dominated by the direct gravitational pull of the circumbinary disc rather than by the interaction with gas in the mini-discs \citep{roedig2012, FranchiniEtAl2021}. 
More massive circumbinary discs drive
faster inward migration as well as stronger eccentricity growth, which can
have important consequences once they reach the GW-dominated regime. 
Moreover, when migration is outward in a non-self gravitating region
of the disc, the radius at which the disc becomes self-gravitating will eventually be reached, at which  point a reversal of the sign of migration occurs, and  orbital decay can resume 
\citep{FranchiniEtAl2021,2021ApJ...918L..15B}.

Finally, \citet{ 2021ApJ...918L..15B}
have investigated, with the aid of a semi-analytical model calibrated on numerical simulations, what would be the net
outcome of the  different concurrent torques once also the effect of stellar
hardening in a typical stellar bulge is included. Even by assuming, very conservatively, that the gas torques resulting from the combination of
circumbinary disc and mini-discs are positive until the binary reaches the self-gravitating
radius, at which point it cancels out,  they found that stellar hardening alone would drive the binary to coalescence always within less than a Gyr,
and for a wide range of SMBH masses ($10^3-10^8$ solar masses). 

In conclusion, currently the results of simulations on various scales
suggest that SMBH pairs evolving in stellar backgrounds are almost certainly going to merge, and likely do that on timescales much smaller
than the age of the Universe. These should correspond to the high end
of the SMBH mass  distribution, which is accessible by Pulsar Timing
Arrays (PTAs). Instead, for lower mass black holes evolving in predominantly gaseous backgrounds, which would be typical LISA sources,
the jury is still out on the efficiency of the binary formation and 
hardening, with wandering SMBHs produced by stalling in various phases
of the orbital evolution being a possible outcome. However, there are
indications that also in this case stellar-driven hardening might come
to rescue, at least if the SMBHs reach parsec-scale separations. Hence
numerical models which can follow both the gas-driven and the stellar-driven torques self-consistently are needed to address this problem further. 

\mghk{The length scales addressed in this section are well accessible with current state-of-the art observatories. The interaction of the black holes with the host galaxy's gas distribution will quite possibly lead to activity of both black holes, though, likely, with different luminosities.} 

\section{Observational evidence for dual active galactic nuclei}
\label{sec:dual-AGN}

Dual AGN are systems of two actively accreting SMBHs, whose host galaxies are in the earliest process of merging.  
They are nominally defined as pairs with separations below ~30 kpc, and SMBHs are not yet gravitationally bound 
(compare above).  
There exist many multi-wavelength techniques to detect dual AGN candidates.  One of the earliest techniques was to use optical spectroscopy to search for double-peaked narrow line emission regions (which can sometimes be spatially resolved; see, e.g., \citealt{Zhou2004, Gerke2007, Comerford2009, Liu2010, Fu2012, Comerford2012, Comerford2013, Barrows2013}).  Dual AGN systems can display two sets of narrow line emission regions, such as [\ion{O}{3}] $\lambda$5007, during the period of the merger when their narrow line regions (NLRs) are well separated in velocity (see below for a discussion of the broad line region).  Here, the separation and width of each peak will depend on parameters such as the distance between the two AGNs. However, double-peaked emission features are known to originate from other processes, such as bipolar outflows and rotating discs  (\citealt{Greene&Ho2005, Rosario2010, Smith2010,Nevin2016}).

\begin{figure}[ht!]
\centering
\includegraphics[height=6cm]{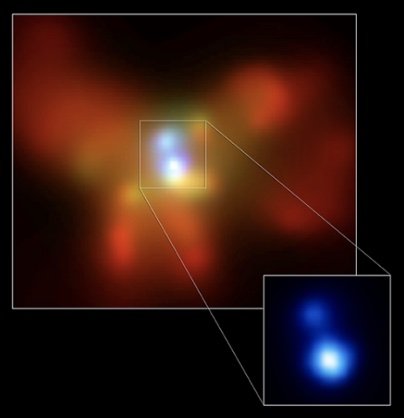}
\includegraphics[height=6cm]{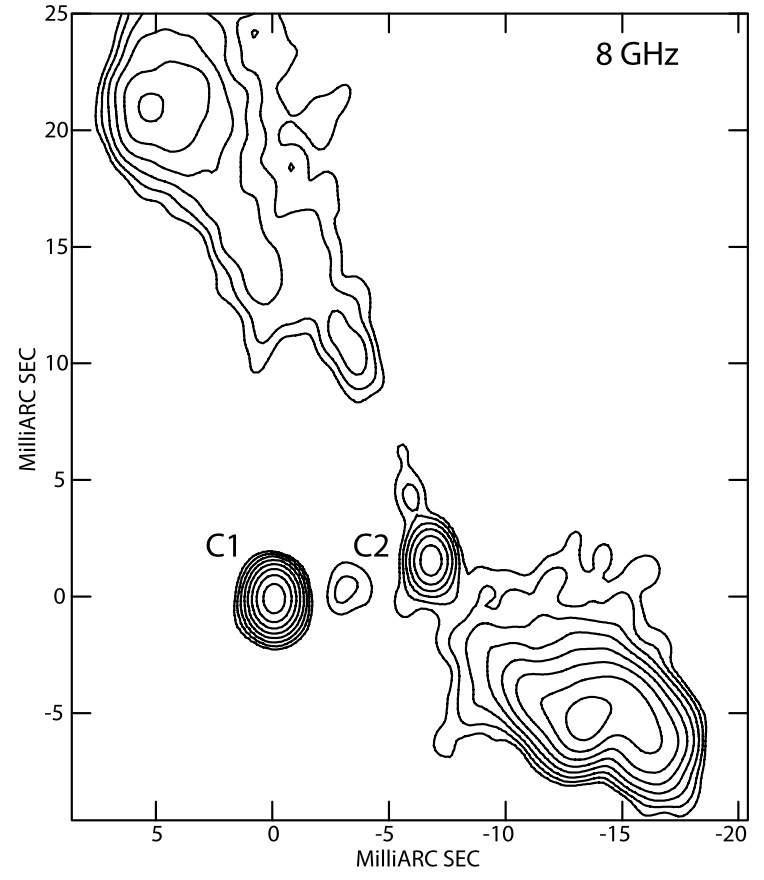}
\caption{Directly resolved dual AGN. {\bf Top:} X-ray maps of the merger remnant NGC~6240 obtained with Chandra. The large map in reddish colours shows the soft 0.5-1.5~keV band. The inset with blueish colours represent the hard 5-8~keV band. The X-ray detection confirms that both nuclei host active AGN and therefore supermassive black holes. Their separation is about 1~kpc in projection. The data was originally published by \citet{Komossa2003}. {\bf Bottom:} Radio map at 8~GHz obtained 
with the Very Long Baseline Array radio telescope of the galaxy 
0402+379 \citep{Rodriguez06}. Two active galactic nuclei are directly detected at 7~pc separation. 
The weaker nucleus hosts a radio jet. Adopted under Creative Commons license, 
\url{https://creativecommons.org/licenses/by/4.0/}.
\label{fig:dualAGN}}
\end{figure}

Confirmation of dual AGN systems requires spatially resolving each individual AGN; and beyond $z>0.05$ high-resolution imaging is necessary. Radio observations can resolve radio-emitting cores on the smallest spatial scales (see \citealt{Rodriguez06, Rosario2010, Tingay&Wayth2011, Fu2011, Fu2015, Deane2014, Gabanyi2014, Wrobel2014a, Wrobel2014b, MullerSanchez2015, Kharb2017}). The most closely separated dual AGN candidate to date, 0402+379, has a projected separation of approximately 7.3 pc and was discovered using VLBA and VLBI \citep[Fig~\ref{fig:dualAGN}]{Rodriguez06, Bansal2017}. However this technique \mghk{is restricted to} \mghkso{only efficient} dual AGN where \textit{both} AGN  are radio bright ($\approx 15\%$ of the AGN population is radio bright; \citealt{Hooper1995}), and individual AGNs can only be differentiated from jet components at radio frequencies if they are compact and have flat or inverted spectral indices (see, e.g., \citealt{Burke-Spolaor2011,Hovatta2014}).  Indeed, this is further complicated by the fact that regions of intense starbursts can mimic both compactness and brightness temperatures of AGNs; thus complementary IR data may be necessary to properly classify the source (see, e.g., \citealt{Varenius2014}).

\subsection{Optical dual active galactic nuclei}
There have been many optical searches for quasar pairs in the high-redshift Universe, where tens of candidates have been identified ($z>1$; e.g., \citealt{Hennawi2006,Myers2008,Hennawi2010,Kayo2012,Eftekharzadeh2017}). Most recently, optical spectroscopy and photometry aided in detecting the highest-z dual AGN candidates \citep[$z>5$,][]{Yue2021, Yue2023}, as well as one of the closest dual AGN candidates \citep{Koss2023}.  
For example, \cite{Stemo2021} analysed a catalogue of 2585 AGN host galaxies observed with the \emph{Hubble Space Telescope} and spanning a redshift range of $0.2 < z < 2.5$. By identifying AGN host galaxies that 
\mghk{have} multiple stellar bulges, they find 204 offset and dual AGN candidates. Recently, new observational techniques that leverage the angular resolution of Gaia have been effective first steps for detecting the dual AGN population at high-z. Varstrometry techniques (see, e.g., \citealt{Shen2019, Hwang2020,Shen2021}) combine variability with astrometry techniques.
\rev{This technique exploits the empirical finding that
the luminosity of any given AGN varies with time. If both SMBH of a binary are active, their fluxes will vary independently of each other. Thus,}
in a just unresolved dual AGN, the uncorrelated flux variability of the AGN will shift the astrometric centre, which leads to \rev{additional} 
noise in the positions of the dual AGN as a function of time. Such observations
have identified a $z>2$ dual AGN \citep{TChen2023} and the Gaia Multi-peak (GMP) method \citep{Mannucci2022} has detected 
\mghk{further} dual AGN candidates at $z>1$ \citep{Ciurlo2020}. 

\subsection{X-ray dual active galactic nuclei}
Optical selection techniques are affected by optical extinction and contamination from star formation, which is especially problematic when observing highly-obscured mergers \citep{Kocevski2015, Koss2016, Ricci2017, Blecha2018, DeRosa2018, Koss2018, Lanzuisi2018, Torres-Alba2018}. As a result, the confirmation of most AGN pairs have been made via X-ray observations as X-rays can more easily penetrate the dense obscuring gas (e.g., NGC 6240; \citealt{Komossa2003}, Fig. ~\ref{fig:dualAGN}), and most studies leverage Chandra's superior angular resolution to discover closely-separated dual AGN \citep{Koss2012, Foord2019, Foord2021, Pfeifle2019}. However, there are less than 50 directly detected pairs of X-ray AGN candidates to date \citep{TChen2022}, as the majority of Chandra-detected dual AGN are restricted to the local universe ($z<0.1$). High-z Chandra survey studies have resulted in non-detections \citep{Sandoval2023} due to the small field of view with high spatial resolution ($<1.5\arcsec$) and sensitivity. 

\subsection{The dual active galactic nucleus fraction}
To date, most predictions of the dual AGN fraction at high-z have been carried out via cosmological simulations \citep[compare Sect.~\ref{sec:smbh-growth} above, and particularly][]{Steinborn2016, Rosas-Guevara2019, Volonteri2022, Chen2023}. However, the observed dual AGN fraction remains relatively unconstrained and has been measured to be higher than cosmological simulations predict 
\citep{Koss2012, Barrows2017}\mghk{, which should not be surprising, given the large uncertainties in these simulations regarding the evolution of supermassive black holes}. A handful of large surveys in the optical regime have yielded constraints on the high-z dual AGN fraction. \cite{Silverman2020} analyse double quasars resolved by the Hyper Suprime-Cam (HSC) 
\mghk{on the} Subaru \mghk{telescope}, where $\sim$100 dual AGN candidates were identified out to z=4.5, and find no evidence for evolution across redshift. \cite{Shen2023} analyse 60 Gaia-resolved double quasars to measure quasar pair statistics at $z<1.5$, and similarly find no evolution across redshift. Most recently \cite{Sandoval2023} measure the X-ray fraction of dual AGN at $2.5<z<3.5$ and find a non-detection among a sample of 66 AGN. This non-detection translates to an upper-limit of the dual AGN fraction of ~4\% at z=3.
\mghkso{, consistent with cosmological simulation predictions. }

\subsection{Future surveys}
In order to significantly expand the number of dual AGN, surveys with exceptional angular resolution, large field-of-views, and high effective areas are necessary. 
The next decade will be the golden age to observe SMBHs through cosmic times and in particular to detect them closer to the redshifts of their formation, and characterise their host galaxies. 
In addition to JWST, the Euclid and Roman telescopes will also image the first galaxies while detecting new accreting AGN and quasars. 
First results have been published very recently with dual and multiple AGN systems with separations from tens of kpc down to less then 1~kpc
\citep{2024MNRAS.531..355U,2025AaA...696A..59P}.
New thirty-meter telescopes such as the 
\rev{Extremely Large Telescope (ELT), the Thirty Meter Telescope (TMT) and the Giant Magellan Telescope (GMT)}
will constrain the assembly of galaxies.
Concept missions scanning the X-ray sky such as Athena \citep{2013arXiv1306.2307N}, {Advanced X-ray Imaging Satellite} \citep[AXIS; ][]{2023SPIE12678E..1ER}, or Lynx \citep{2018arXiv180909642T} could revolutionise our understanding of AGN activity before and during galaxy mergers by increasing substantially the number of X-ray-observed dual AGN all the way to cosmic dawn.
For example, the probe-class AXIS ({with baseline angular resolution of} $1.5''$ {half-power-diameter} on-axis and $\sim 1.75''$ FoV-average) and the 
\rev{potential} future flagship Lynx ($\sim 0.5''$ FoV-average) aim at unprecedented sub-arc second angular resolution to uncover dual AGN with a separation down to a few kpc up to $z=10$ \citep[][for a review]{2019NewAR..8601525D}. On a shorter timescale from now, the Roman telescope with its near-IR wavelength large-sky coverage could probe dual AGN with luminosities down to $L_{\rm bol}\geqslant 10^{42}\, \rm erg/s$ beyond $z\geqslant 1$ with an angular resolution of $\sim 0.13''$ \citep[][for two recent CSS white papers]{2023arXiv230614990H,Shen2023}.

\section{Observational evidence for sub-parsec binaries}
\label{sec:lines}

\mghk{Going down to sub-parsec separations we are}
approaching the distance where gravitational radiation drives their inspiral. Despite decades of observational effort, we do not yet have a sample of confirmed sub-parsec binary SMBHs. The closest pair that is often-cited as our best-case tight binary is at a separation of 7 pc and 
was already discussed above. Direct imaging of pairs rapidly grows prohibitive with currently available angular resolution. The most comprehensive search for black hole pairs at high angular resolution yielded no additional candidates \citep{BurkeSpolaor:2011}. 

\subsection{Optical searches}
Short of direct imaging, there are a number of additional techniques that have been attempted to uncover sub-pc pairs of (accreting) SMBHs. One, which we will not discuss in detail here, predicts quasi-periodic temporal variability linked to the orbital period of the two black holes.
\rev{Here, the assumption is, based on hydrodynamic simulations, that the accretion rate onto the black holes, and thus their luminosities, would be modulated periodically with a period related to the orbital period of the binary \citep[e.g.,][]{2016MNRAS.463.2145C}. A}
number of papers have searched for this signature and identified candidate sub-pc binaries \citep[e.g.,][]{2015Natur.518...74G,2016ApJ...833....6L,2016MNRAS.463.2145C}. However, because the survey lengths sample only $\sim 1.5-3$ orbital cycles, it can be easy to mistake the red-noise variability that characterises AGN lightcurves from true orbital variability \citep{Vaughan:2016}. Furthermore, the number of pairs is higher than would be expected from upper-limits on the gravitational-wave background \citep[e.g.,][]{Sesana:2018}. 

The other approach looks for velocity shifts of broad optical emission lines from photo-ionised clouds in the vicinity of the black holes, the so-called broad line region, similar to spectroscopic binary stars \citep[see nice review in][]{Runnoe:2017}. For many years, a small subset of AGN have been known to have two velocity peaks \citep[e.g.,][]{Oke:1987,Eracleous:1994}. One possible explanation for these two velocity components is a binary system \citep[e.g.,][]{Gaskell:1996}, although far more likely in most cases based on the velocity structure and variability of the lines, is that we are seeing disc emission in these cases \citep[e.g.,][]{Eracleous:1997,Gezari:2007}. It is worth stating the caveat that this spectroscopic technique will only be sensitive to spatial scales larger than the size of the broad-line region ($\sim 0.01$~pc), when the two black holes can each sustain their own broad-line emission. In contrast, the quasi-periodic variability will in principle be sensitive to much more tightly bound objects with orbital periods of days to months.

With the advent of the Sloan Digital Sky Survey, objects with dramatic velocity offsets between the broad and narrow lines emerged \citep[e.g.,][]{Komossa:2008,Shields:2009,Boroson:2009,Chornock:2010}. \citet{Boroson:2010} initiated a systematic search through the SDSS for objects with large velocity shifts between broad and narrow lines, and \citet{Eracleous:2012} began a long-term monitoring campaign of 88 candidate binaries from this sample. Thus far, after more than a decade of repeated spectroscopic observations, three of the 88 are still viable targets for binarity based on the radial velocity curves measured to date \citep{Runnoe:2017}. Indeed, a challenge with selecting objects showing 1,000 km/s velocity offsets between the broad and narrow emission lines is that if an orbiting black hole is implicated, then it will be at apocentre, and we will have to wait for a long time to measure changes.

A complementary spectroscopic approach is to simply search for radial velocity shifts in all AGN with multi-epoch spectroscopy, with the idea of observing radial velocity shifts corresponding to orbital motion \citep{Ju:2013,Shen:2013}. The main challenge with this approach, which otherwise should yield reliable constraints on the sub-pc binary fraction, is that velocity flickering in the broad-line region is a major source of noise. \citet{Ju:2013} identify candidate binary black holes from $\sim 1-10$~years of spectroscopic monitoring by SDSS, but an additional epoch of follow-up for these targets shows that in all cases the velocities are varying in a stochastic way. Specifically, \citet{Wang:2017} present a third epoch of spectroscopy for the seven candidates, and in no case is there a coherent radial velocity curve. Thus, this velocity flickering in single broad-line regions proves to be the primary contaminant to blind spectroscopic searches for binaries signalled by radial velocity shifts.  

Real progress is on the horizon as time-domain imaging and spectroscopy become increasingly routine. Already SDSS-V have started working to characterize the temporal velocity structure of typical AGN from multi-epoch spectroscopy \citep[][]{Fries:2023}. If we can characterize the power as a function of velocity coming from standard AGN, the hope is that we can filter out the rare systems with coherent radial velocity shifts. At the same time, the Rubin Observatory will soon launch the Legacy Survey of Space and Time \rev{\citep[LSST,][]{2019ApJ...873..111I}}, which may finally realize the promise of quasi-periodic searches for AGN. \rev{With its high cadence of large-area deep optical suveys, LSST would allow studies of archival lightcurves of binary SMBH after detection of the gravitational waves with LISA \citep{2024MNRAS.533.3164X} or identify the periodic signal from milli-parsec separation binary SMBH with large mass ratio \citep{2024AA...691A.250C}. }
\mghkso{Finally, with a slightly longer horizon, there is a the hope that next-generation VLBI (perhaps from space) may allow for the micro-arcsecond imaging required to directly view sub-pc pairs of black holes \citep[e.g.,][]{Pesce:2021}.}

\subsection{Binary black hole jets in high-resolution radio imaging}
\label{sec:VLBI}

So-called astrometric binaries might reveal themselves by precessing parsec-scale jets. Precession of AGN jets can be caused by either a misaligned binary SMBH system where for an SMBH with low spin, the jet direction is given by the disc angular momentum 
which is perturbed by the secondary \citep{Abraham2018} or the Lense-Thirring (LT) effect \citep[misaligned disc around single black hole][]{Lense18,Thirring18}. 
Another important effect, sometimes not subsumed under precession is a jet from an orbiting black hole \citep{Fendt:2022}, a case that is clearly observed in the source 0402+379 (Fig.~\ref{fig:dualAGN}). The latter will cause a mirror-symmetric jet direction variability rather than point symmetry as for the other processes. However, as due to relativistic beaming, often only one jet is observed, orbital precession can be difficult to distinguish from the former two mechanisms.
Precession is observed as periodic oscillation of the main jet ridge line and/or periodic patterns detectable in the light curve \citep[e.g.,][]{abraham_carrara98,Caproni04a,Caproni04b,Britzen10,Caproni12,Britzen17,Britzen19,Britzen21}.

\begin{figure*}
\centering
\includegraphics[height=8cm]{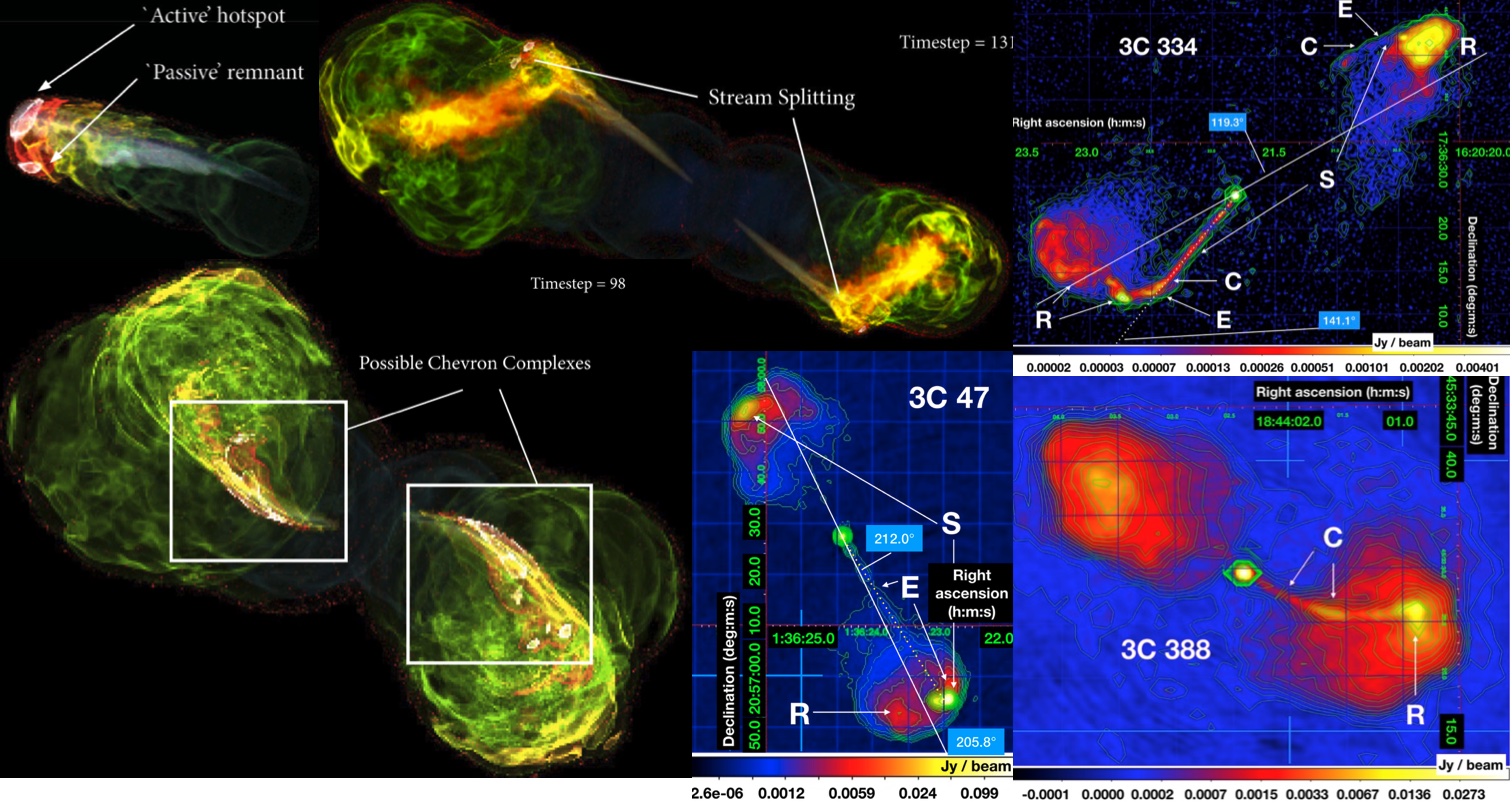}
\caption{3D hydrodynamics simulations of precessing jets from
\citet[3 examples towards the left with uniform black background]{Horton23} and Very Large Array radio maps at 5 and 8 GHz of radio galaxies and quasars with jets that show signatures expected from a long-term driven precession, such as the relativistic geodetic spin precession in a binary SMBH (3 examples towards the right, each with their 3C ctalogue number). The three simulations had different precession parameters and form a variety of interactions with the lobe boundary and hotspot structures. Such features can be compared to observed radio galaxies and used to understand the properties of the driving system.
For the observed radio maps towards the right, the projected sizes on the sky range between 70 and 330~kpc. Letters mark the occurrence of precession features discussed in more detail in the Sect.~\ref{sec:prec-jets}: 
E -- jet detected towards the edge, rather than the middle of the lobe, S -- S-symmetry, C -- jet curvature, R --
ring-like, extended or multiple hotspots. 
\rev{While the simulations on the left do not reproduce any particular system on the right, they do demonstrate that precessing jets may be curved, misaligned with the lobes and have multiple hotspots.}
Images adopted from 
\citet{Krause19} who find strong indication for jet precession in 24 out of 33 powerful radio galaxies from a complete sample.\label{fig:jetsims}}
\end{figure*}

OJ~287 is considered to be one of the most promising candidates for harbouring a binary SMBH \citep[e.g.,][]{Sillanpaa88,Valtonen09,Britzen18}, based on its characteristic periodicities in the optical light curve domain ($\sim$11 years). The most popular binary SMBH-model for OJ~287  \citep[e.g.,][]{Lehto96, Valtonen09} explains the substructure inside the major outbursts with a model in which a smaller black hole crosses the accretion disc of a larger black hole during the binary orbit of the black holes about each other.

Disturbances of the accretion disc should, however, also lead to \mghk{corresponding} disturbances of the radio jet, \mghk{if the jet was produced by the larger black hole around which the accretion disc would be found,} which are not observed \citep{Britzen18}. Quite in contrast, according to \citet{Britzen18}, the jet reveals the signature of continued precession on a timescale of 22-23 years. As the authors demonstrate, also the radio light curve variability can be explained by the precession model. A binary SMBH or LT-precession seems to
be required to explain the time scale of the precessing
motion. Such a scenario intrinsically connects the radio and optical periods (radio period being
twice the optical period) when viewed close to the rotation axis. \mghk{The scenario would hence involve two accretion disc passages during one orbit of the smaller black hole, causing the optical brightenings \citep[also compare][for more recent analysis of the binary SMBH system]{Komossa2023}. The orbital motion would drive the disc and hence jet precession.
Other explanations of the precession mechanism are possible. For example, the smaller black hole could also produce the jet. The precession would in this picture be caused by a largely ballistic motion, always driven away from the centre of the orbit and hence also produce one jet precession period for each full orbit of the smaller SMBH.} 
\citet{Britzen2023} most recently showed, that also the Spectral Energy Distribution (SED) can be directly related to the jet’s precession phase in OJ~287.
\mghk{This is due to the directional dependence of the Doppler effect. During the precession phase when the jet points towards the observer, the jet emission will be boosted, whereas the luminosity drops when the jet points away from the observer. This produces a periodic variation of the spectral components emitted by the jet.}
\citet{Zhao22} provided further support for the precession scenario in OJ~287. The first \mghk{Global mm-VLBI Array} (GMVA) and \mghk{Atacama Large Millimeter/submillimeter Array} (ALMA) observations  observed at 3.5 mm show a radio structure resembling a precessing jet in projection. 
The modulating power of precession might more generally govern AGN structural as well as lightcurve variability in the cm-regime \citep{Britzen2023}. 


\subsection{Clues from precessing large-scale jets}
\label{sec:prec-jets}
Jets from supermassive black holes 
\mghkso{often drive powerful jets that} 
can extend for hundreds of kiloparsecs away from the galactic nucleus \citep{Hardcastle20}. 
When two or more SMBHs form a stable system in a single galaxy, periodic disturbances in 
the jet path are expected to arise from geodetic precession of the black hole spins around 
the total angular momentum vector, which is dominated by the orbital angular momentum 
\rev{\citep{1980Natur.287..307B}}. Geodetic precession is particularly useful for kpc-scale jet observations, as, even for sub-parsec SMBH binaries, the geodetic precession frequency can be of the order of millions of years, resulting in long-lasting, predictable morphological characteristics which can be used to identify suitable candidate galaxies for close binaries \rev{\citep{Krause19}}. 

\begin{figure*}
\centering
\includegraphics[height=8cm]{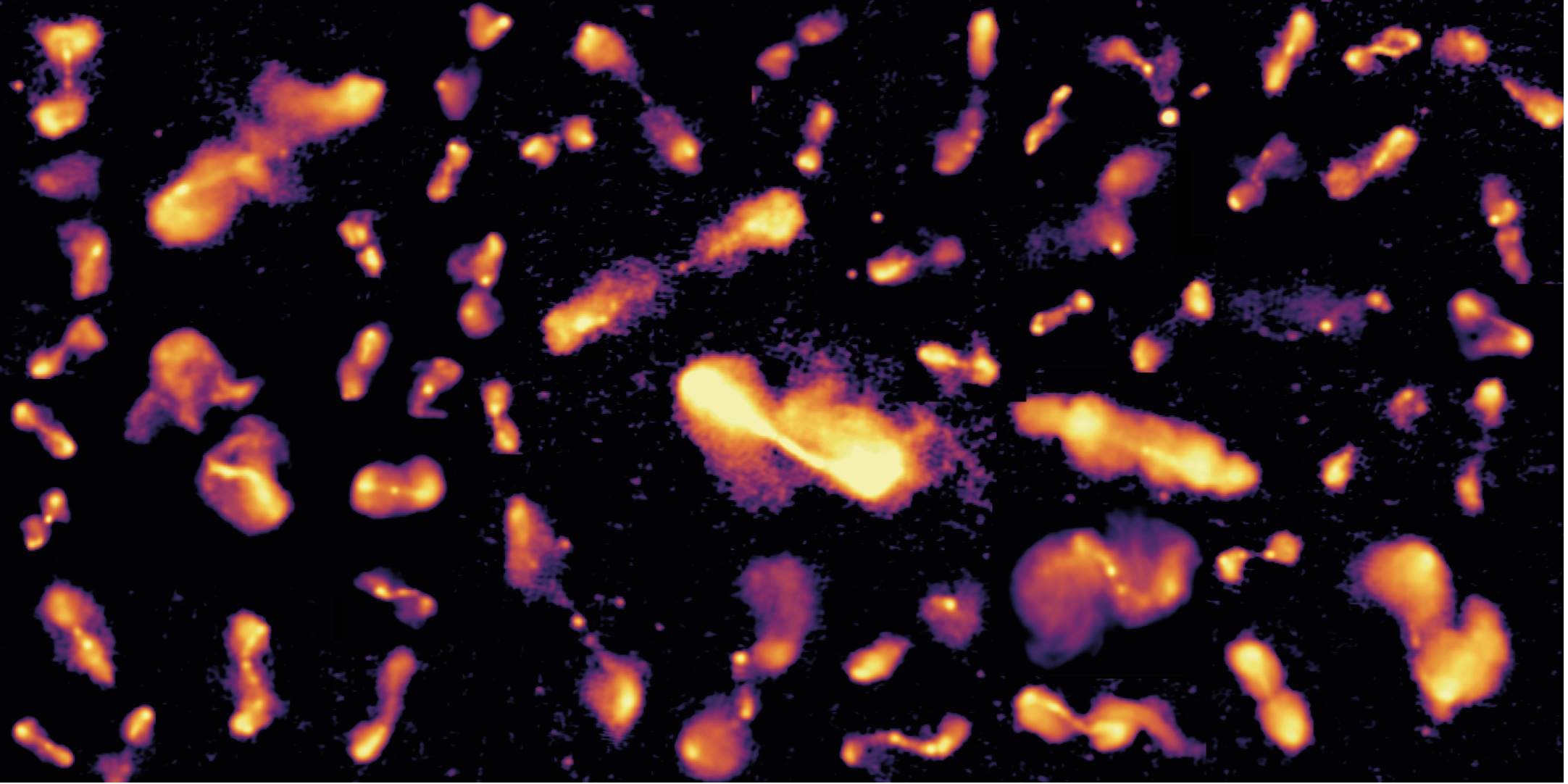}
\caption{Likely precessing 100~kpc-scale jets and therefore binary black hole candidates from LOFAR. Adopted from \citet{2025arXiv250418518H}. The image is a montage of representative precessing sources \rev{selected in \citet{2025arXiv250418518H} to demonstrate the presence of precession indicators E, C and R} discussed in Sect.~\ref{sec:prec-jets}. There is sometimes a striking similarity with the simulated images, compare, e.g., Fig.~\ref{fig:jetsims}, left.
\label{fig:lofarjets}}
\end{figure*}

Four morphological characteristics can be used to identify precession \citep[Fig.~\ref{fig:jetsims}, right]{Krause19}: jets that are misaligned or at the edge of lobes (E); jets that show curvature (C); `S'-shaped symmetry between the two sides of the lobes (S), and ring-like, multiple or highly complex hotspot structures (R). All four structures can be reproduced easily with high-resolution numerical simulations \rev{(as demonstrated in the simulations on the left of Fig.~\ref{fig:jetsims}) }. \cite{Horton20b} show that one or more of the E, C and S indicators are almost always visible in precessing jets\mghkso{(although can be obscured by certain viewing angles or projection effects)}, while the prevalence of false positives in non-precessing jets is low. 
The misalignment effect is a simple consequence of geometry. The velocity in the jet is usually much higher than the expansion speed of the lobe structure the jet lives in. 
This can be compared to the motion of the second hand of a wall clock. Take a photo at any given instance, and it will be very likely that it is not aligned with the 12-o-clock position. However, a non-precessing jet will have a strong tendency to point towards the symmetry axis of the lobe, which emerges roughly symmetrically around the jet as it propagates. This would correspond to a second hand fixed to the 12-o-clock position in the picture. Particularly for precessing jets that are aligned close to the line of sight, one expects to see curvature due to the change of ejection direction as a function of ejection time \citep{Krause19,Giri:2022}. Curvature is also expected where a precessing jet interacts with the lobe boundary. Indeed, this is particularly frequently seen for broad line radio galaxies or quasars, where the unobscured view towards the broad emission line regions around the AGN confirms the aligned jet orientation. Observing S-symmetry is significantly inhibited by the relativistic Doppler de-boosting of the counter jet. Hence, the counter hotspot will be the main piece of evidence for S-symmetry.
Multiple hotspots were explored in \cite{Horton23} since hydrodynamic processes can produce multiple hotspots by several different mechanisms (also compare Fig.~\ref{fig:jetsims}, right). Some of these -- such as hotspot complexes and dual jet paths -- seem largely confined to precessing jets while others, such as hotspot splitting, can occur in non-precessing jets too. Backflow can push remnant hotspots into parts of the lobe where there has never been a primary flow. Overall multiple hotspots are confirmed as a signature of precessing jets.
\mghkso{, meaning that more studies are required to understand multiple hotspot dynamics before using them as a diagnostic for supermassive binary systems. }
Precession may explain a number of complex radio structures, including re-starting, re-orienting, X-shaped and Z-shaped sources and possibly Odd Radio circles \citep{Horton20b,Nolting:2023,Shabala:2024}.

\mghkso{Regardless, multiple hotspots play a critical role in the identification of supermassive binaries.} 
While precession has a clear impact on the jet path, the latter may not just be a simple ballistic trajectory of plasmons ejected from the vicinity of one SMBH.
\cite{Horton20} used a Bayesian approach to fit a, relativistic, ballistic, precessing jet model \citep{Gower82} to the well-studied radio galaxy Cygnus A. Cygnus A shows the presence of all four indicators to some degree. While it was possible to obtain probabilistic constraints on the binary separation using our approach, this was only possible for the case where at least one set of terminal hotspots were included at the end of the jet path. Without them, no fit could be found. This highlights the need to study 
jet paths including terminal hotspots with 3D hydrodynamical
simulations.
\mghkso{the formation mechanisms and decay times of multiple hotspots a lot more clearly in order to understand their possible role in precession.} \cite{Horton23} found in such simulations that some mechanisms -- such as large complexes with a high number of short-lived multiple hotspots -- are dependent on more extreme forms of precession and if present may reveal additional information about the driving system. 

Observed precession in powerful kpc-jets is likely due to geodetic precession. This is because the re-alignment timescale for LT-precession is relatively short, of the order of the precession period. Since the source ages are typically much greater than the likely precession period, disc-induced precession would require a co-incidental event that would have produced the disc misalignment just at the right time for the observation. The high incidence of kpc-scale precession at least in powerful sources \citep[24 of 33 sources in the complete sample in][]{Krause19} argues for a frequent occurrence of geodetic precession and hence, close binary SMBH. \citet{2025arXiv250418518H} found 1138 strong precessing jet candidates showing at least two precession indicatiors with the LOFAR Two-metre Sky Survey (LoTSS DR2), which will help to answer questions about host galaxy characteristics and the origins of different forms of precession (Fig.~\ref{fig:lofarjets}). There are thus good prospects to address the prevalence of binary SMBH in specific types of galaxies. 

In order to produce well-visible precession features, the precession period must be comparable to the age of the source. If it was much shorter, the jet would become unstable and would just produce fuzzy radio lobes. If it was much longer, the radio lobes have time to adjust and the 
radio source looks symmetrical again. The ages of well-resolved LOFAR radio sources are generally of the order of $10^8$~yr
\citep{Hardcastle19}. For geodetic precession with period $P_\mathrm{gp,8} \times 10^8\,$yr in a supermassive binary of mass 
$M_9 \times 10^9 M_\odot$ and of any mass ratio, the separation in parsec is limited by 
\rev{\citep{Krause19}: 
\begin{equation} 
d< 1 \,\mathrm{pc}\,P_\mathrm{gp,8}^{2/5} M_9^{3/5}\, .
\end{equation} 
Probably the best case available for this method in the literature, which is Hydra~A. For this radio source, dedicated hydrodynamic 
simulations have constrained the precession period to $10^6$~yr, with an accuracy of 50\% \citep{2016MNRAS.458..802N}. The SMBH mass has been determined to $(5\pm4)\times 10^8 M_\odot$ \citep{2006ApJ...652..216R}. The long-term approximate stability of the precession is evident from the large-scale radio structure. Therefore, the precession is likely caused by geodetic precession in an SMBH binary with a separation less than 0.2~pc
\citep[compare][]{Krause19}.
}
Hence, these sources might contain the tightest electromagnetic binary SMBH candidates so far.


\section{\rev{Spin measurements} of supermassive black holes \rev{from} X-ray reflection}
\label{sec:spins}
\rev{SMBH spin measurements can provide indirect evidence for binary SMBH mergers, and thus the existence of SMBH binaries.
This is because coherent gas accretion tends to produce maximally spinning black holes, whereas SMBH mergers tend to produce more slowly spinning SMBH remnants. As we detail in this section, cosmological simulations have started tracking the SMBH
spin evolution and including the effects of binary SMBH mergers. The observed distribution of SMBH spins over mass scales suggests higher
spins at low SMBH masses and lower spins at very high SMBH masses. Informed by
predictions from cosmological simulations, the observed distribution seems best
explained by accretion (mergers) being important for low-mass (high-mass) SMBHs.}

The magnitude of the black hole spin can be probed directly via X-ray emission from the innermost regions of the accretion disc.  
X-rays from the vicinity of the SMBH are reflected from a relatively cold accretion disc. The reflection produces an emission line that is redshifted gravitationally, such that from the distortion of the line profile one can infer the properties of the source of gravity, i.e., the SMBH, 
including the black hole spin magnitude.

\begin{figure}[ht!]
\includegraphics[width=\textwidth]{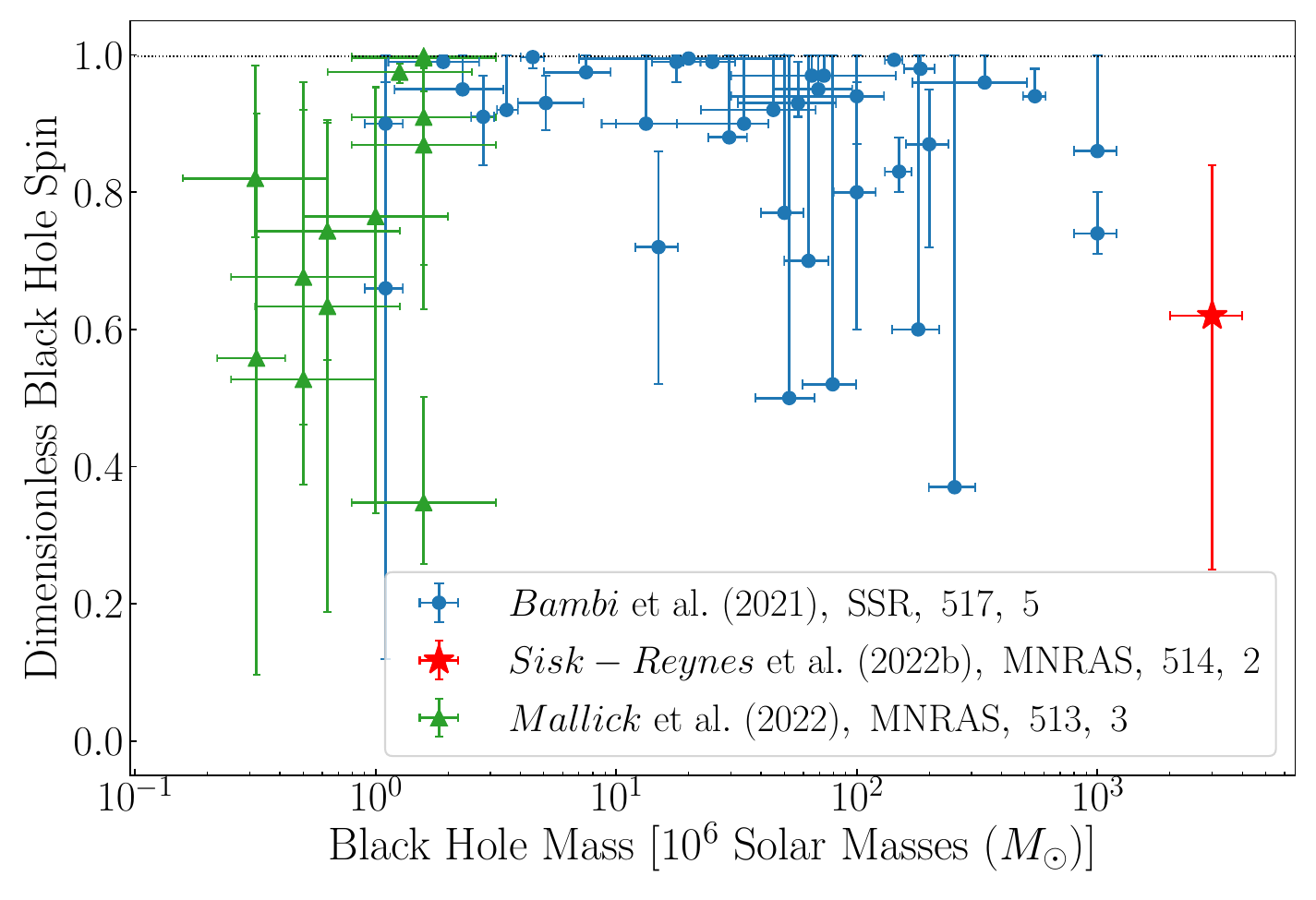}
\caption{Spin \rev{magnitude estimated from X-ray reflection} as a function of mass for 51 SMBHs \rev{in AGN, compiled} from the literature.
\rev{The figure is} adapted from the \citet{Reynolds2021} and \citet{Bambi2021} spin reviews with the inclusion of: the 13 SMBH spin estimates in low-mass AGN from \citet{Mallick2022} (in green) and recent spin constraints \rev{for the SMBHs in} H~1821+643 \citep{Sisk-Reynes2022} and ESO~033-G002 \citep{Walton2021}.~Note some of the estimates from \citet{Bambi2021} show a well-defined lower spin bound.~We omit the following objects: IRAS~1339+2438 and 4C~74.26 \citep[see Sec.~6 of][]{Sisk-Reynes2022}, Tons~180, and the sample of type-1 AGN from \citet{Mallick2025_unpublished}. Error bars in spin/mass correspond to statistical uncertainties at the 68/90 per cent level, respectively.
\rev{Comparing with predictions from cosmological models, the lower spins at the highest masses is best explained by binary SMBH merging. More measurements are needed to corroborate this result.}
\label{fig:spinconstraints_adapted}}
\end{figure}

Over recent decades, the application of relativistic reflection models to X-ray spectra have provided spin constraints for over~50 SMBHs 
across mass scales spanning $10^5 - 10^{10} M_\odot$ (Fig. ~\ref{fig:spinconstraints_adapted}). In addition to being one of the two fundamental properties of astrophysical black holes, the spin acts as a fossil record of their growth history. Specifically, the black hole spin provides an indicator of the relative contributions of coherent accretion, incoherent accretion, and SMBH mergers in powering the recent growth of SMBHs. This therefore suggests that tracing the magnitude of the black hole spin over cosmic time and mass scale may provide indirect evidence for the existence of SMBH binaries.

The power-law X-ray component that usually dominates the high-energy emission in AGN arises from direct emission of the X-ray corona, a magnetically-heated cloud of relativistic electrons which steadily illuminates the optically-thick accretion disc. The irradiating hard X-ray continuum from the corona is reprocessed in the surface layers of the disc and produces the ``X-ray reflection spectrum''. This reflection spectrum contains a rich set of fluorescence and radiative recombination lines, with the most prominent being the K-shell lines of iron with rest-frame energy of 6.40--6.97~keV, depending on the ionization state of the iron in the disc. 

Reflection features from the inner disc are broadened and skewed due to the Doppler shifts of the mildly relativistic orbital motions in the disc, as well as gravitational redshift. Based on the dependence of the innermost stable circular orbit (ISCO) on the value of the black hole spin, \citep[Eqs.~2-4 of][]{Reynolds2021}, reflection models can then predict the properties of the resulting irradiation profile if the inner disc radius is fixed at the ISCO \cite[and thereby parametrized by BH spin; see][]{GarciaDauser2014, DauserGarcia2014}. See \cite{Reynolds2021} for a detailed discussion of the assumptions that underlie this method. We also note the discussion in \citet[][]{Barret2019} which explores the extent to which relativistic reflection signatures can be degenerate with absorption from fast ionized winds from the AGN.

The spin results shown \rev{in} Figure~\ref{fig:spinconstraints_adapted} suggest an interesting, \rev{tentative} dependence of the observed SMBH spins in the distribution as a function of SMBH mass. The population is dominated by rapidly spinning SMBHs in the mass range $10^{6-7.5}\,M_\odot$. However, a more slowly spinning population may be emerging both above and below this mass range. Interestingly, this pattern seems to naturally emerge from some state-of-the-art models of SMBH evolution. \cite{Bustamante2019} incorporate a subgrid model \rev{to track the spin evolution of individual SMBHs} into an extended version of the IllustrisTNG cosmological simulation of cosmic structure formation. \citet{Bustamante2019} find that black holes of masses $<10^8\,M_\odot$ tend to grow via coherent accretion, since the typical accretion events can torque the (\rev{lower-mass}) black holes into alignment (also compare Sect.~\ref{sec:discs-breaking-and-spin}). Hence these systems are readily spun up to close to maximum. \rev{More massive} SMBHs ($> 10^8M_\odot$), on the other hand, cannot be torqued to alignment by the accretion flow and hence tend to grow in an incoherent mode \citep{King2008}. The averaging of the incoming angular \rev{momentum} tends to spin isolated, inividual black holes down. Furthermore, other cosmological models find that, in general, major mergers with other SMBHs are more important for more massive SMBHs compared to lower-mass SMBHs, and the result of a major merger between two SMBHs is typically a modestly spinning SMBH \citep{Volonteri2005,Sesana2014,2024A&A...685A..92S}. The predictions from \citet{Bustamante2019} are, however, in tension with those of the NEWHORIZONs simulation \citep{beckmann2025}, which includes a prescription to track the black hole spin evolution due \rev{to} jet-driven spin-down episodes. Nevertheless, the importance of  such episodes was also recently assessed by \citep{Ricarte2025} from a semi-analytical model, concluding that the population shown in Figure~\ref{fig:spinconstraints_adapted} could point to jet-driven spin-down episodes for black holes which are initially growing at very low and very high specific accretion rates.

Finally, we note that there is an expectation that SMBHs with high-to-maximal spin are overrepresented in Fig.~\ref{fig:spinconstraints_adapted}~due to spin-dependence of the radiative efficiency \citep{Brenneman2011}.~That is, for a given accretion rate, \rev{SMBHs with higher spins} will be brighter \rev{and more easily identifiable} as potential good candidates from flux-limited AGN samples \citep[see Fig.~3 of][]{Reynolds2019}.~Moreover, other systematic effects must be accounted for, \rev{if the observed population in Figure~\ref{fig:spinconstraints_adapted} is used to compare with predictions
from cosmological simulations}, including the effects of disc thickness for \rev{SMBHs} with sufficiently high Eddington fractions \citep[][]{Taylor2018}.

\section{Accretion physics: warping, breaking and spin evolution}
\label{sec:discs-breaking-and-spin}

The properties of gravitational waves and BH merger remnants are determined by the pre-merger BH properties--including masses and spins--as well as the orbital properties of the binary. Accretion prior to the merger can significantly alter BH spin, either via direct accretion of angular momentum or via gravitational torques between the BH and its accretion disc. For binaries in gas-rich environments, each BH's accretion disc can be fed by gas streams from a larger circumbinary disc. The most general solution for this allows the BH spins, binary orbit, accretion discs and circumbinary disc to be freely oriented. Torques exerted on the binary by the gas not only drive binary evolution by modifying its angular momentum magnitude but also its orientation \citep[e.g.,][]{NixonEtAl2011}. Further, if the accretion disc and BH spin are misaligned, Lense-Thirring precession, in combination with accretion disc viscosity leads to a warped disc where the disc orientation changes with radius. This process, known as the Bardeen-Petterson effect \citep{1975ApJ...195L..65B}, torques the BH and (counter-) aligns the BH spin and accretion disc from the inner edge outwards \citep[e.g.,][]{KingEtAl2005}. These processes strongly depend on the geometry, physical properties and thermodynamics of the system. How fast the BH spins align in these gas-rich binaries is thus a complex question.

Numerous simulation works have studied the evolution of BH binaries in circumbinary discs \citep[e.g.,][and discussion in Section~\ref{sec:final-pc}]{CuadraEtAl2009, roedig2012, DuffellEtAl2019, MunozEtAl2019, MunozEtAl2020, FranchiniEtAl2021, BourneEtAl2023, SiwekEtAl2023}, and some additionally resolve streams and discs around each BH in order to study their net affect on torquing the binary \citep[e.g.,][]{roedig2012, DuffellEtAl2019, MunozEtAl2019, BourneEtAl2023}. When BH growth is considered, preferential accretion onto the secondary is nearly universally observed \citep{FarrisEtAl2014, DuffellEtAl2019, MunozEtAl2020, SiwekEtAl2023b}, although this is possibly sensitive to gas thermodynamics and accretion modelling \citep[see e.g.,][]{roedig2012}. 

\begin{figure}[ht!]
\includegraphics[width=\textwidth]{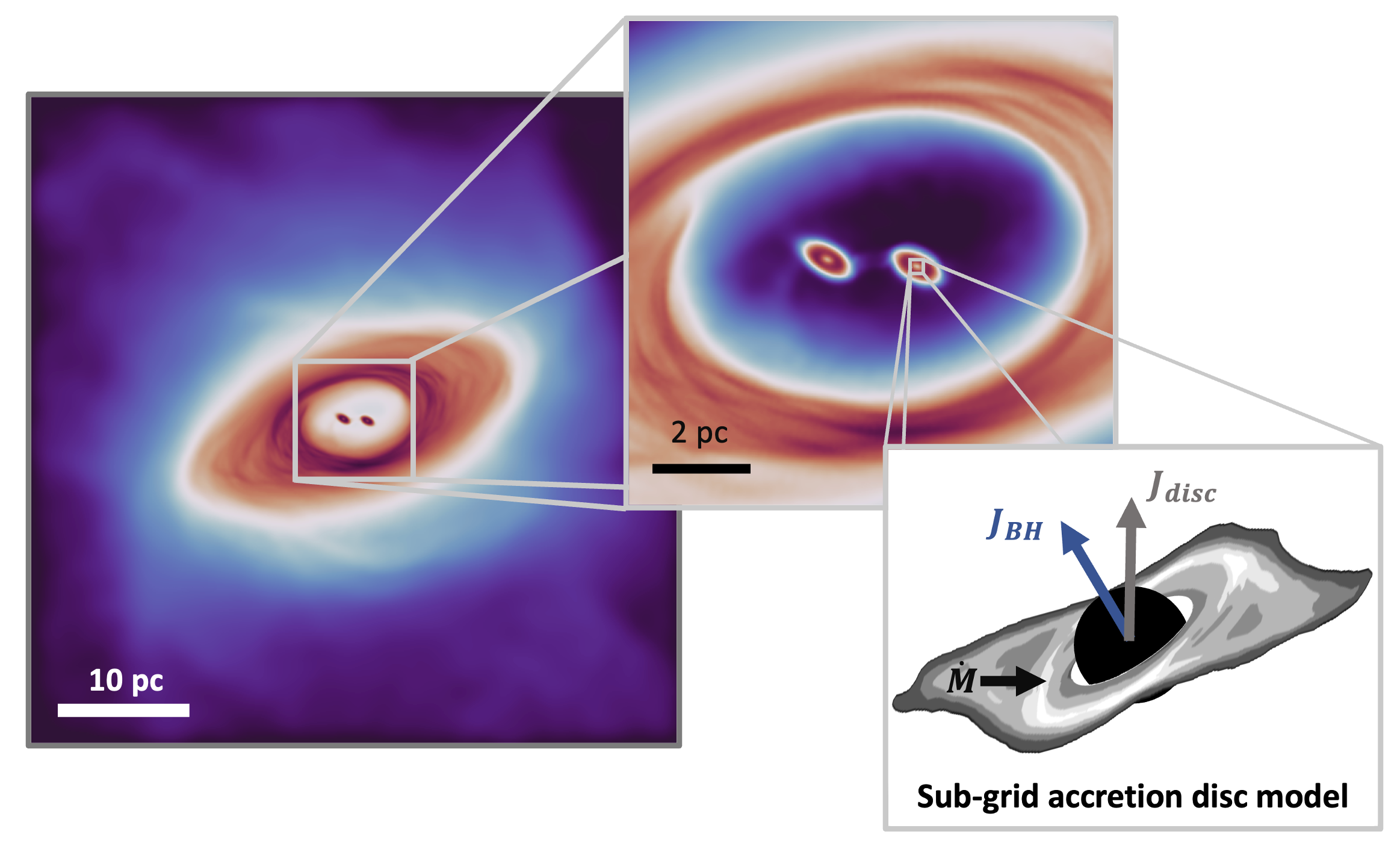}
\caption{Example circumbinary disc simulation in which the binary and disc are initially misaligned \citep[full results presented in][]{BourneEtAl2023}. The circumbinary disc, ``minidiscs'' and binary evolution and interactions are resolved by the simulation, while the accretion disc and BH are modelled using the sub-grid accretion disc prescription of \citet{FiacconiEtAl2018} to capture both the BH mass and spin evolution. (Image Credit: Martin A. Bourne, \href{https://creativecommons.org/licenses/by/4.0/}{CC By 4.0})}
\label{fig:martin_cbd_scales}
\end{figure}

In addition to affecting the inward migration of the two SMBHs, the circumbinary disc can affect the alignment of the SMBH spins.  This is of central importance to the question of whether ``superkick'' configurations can be created that would eject the remnant SMBH from the galaxy post-merger.  This problem was first examined by \cite{2007ApJ...661L.147B} who, using timescale arguments, suggested that SMBH spins would be readily aligned in gas-rich mergers thus avoiding superkicks.  This was further developed by \citet{GerosaEtAl2015} who developed a semi-analytic model to predict BH-accretion disc\mghk{-driven BH-spin} alignment timescales in circumbinary discs assuming the binary, accretion discs and circumbinary disc are co-planar. They found that for high mass ratio binaries BH spins should align before the BHs merge, however, given their model assumes preferential accretion onto the secondary, reduced accretion onto the primary in systems with mass ratios $\lesssim0.2$ prevents it from aligning; leading to misaligned spins at merger and hence high expected recoil velocities \citep{GonzalezEtAl2007, LoustoEtAl2011}. \citet{BourneEtAl2023} performed high-resolution simulations of binaries in gas-rich circumbinary discs for different setups that vary the binary mass ratio, eccentricity and inclination (including retrograde binaries), with an example setup shown in Figure~\ref{fig:martin_cbd_scales}. BH accretion and spin evolution due to accretion and the Bardeen-Petterson effect are modelled using the sub-grid model of \citet{FiacconiEtAl2018}. Therefore, it was possible to not only explore the evolution of the binary, which due to gas torques was found to shrink for all set-ups considered, but for the first time also track the BH spin evolution. \citet{BourneEtAl2023} found BH-accretion disc alignment timescales to generally be shorter than binary inspiral timescales even for low mass ratios, largely due to a lack of preferential accretion onto secondaries in their simulations, which is attributed to i) using beta-cooling (where the cooling time is proportional to the local orbital time) combined with an adiabatic (opposed to locally-isothermal) equation of state, and ii) calculating net inflow rates on the scale of the BH accretion disc as opposed to using a sink particle prescription. However, cases in which the binary and circumbinary disc are initially misaligned (as possibly observed in OJ~287, compare Sect.~\ref{sec:VLBI}), show that the discs around each BH do not necessarily rapidly align with the circumbinary disc and sometimes precess. 
Hence, unless the system can evolve unperturbed for long timescales and processes such as disc breaking (see below) do not occur, even for rapid BH-accretion disc alignment, BH spins may remain misaligned at the time of merging.
The situation may be even worse for other accretion disc geometries, such as thick or truncated discs, where alignment timescales are slower and precession may become important \citep{KoudmaniEtAl2024}\rn{.}
 
\citet{2020MNRAS.496.3060G} built on their earlier work by considering the impact of the binary companion on the BH spin and introduced a semi-analytic approach that solved the non-linear long-term dynamics of warped accretion discs in the presence of a misaligned binary. For a subset of the parameter space their results showed the accretion disc will become warped and align the BH spin on less than 
\rev{10 times the analytic prediction for the disc and BH to realign \citep{ScheuerFeiler96}. This analytic prediction is referred to as the 'warp alignment timescale' and is typically much faster than the accretion timescale but slower than the viscous timescale for typical AGN parameters 
\citep[e.g.][]{2020MNRAS.496.3060G}.}
However, for accretion discs with either steep warp profiles or low viscosities, the semi-analytic model of \citet{2020MNRAS.496.3060G} did not show convergence. For a given set of binary and accretion disc parameters, they named the inclination that marked the transition into a region of no solutions as the ‘critical obliquity’.  
 
\begin{figure}[ht!]
\includegraphics[width=\textwidth]{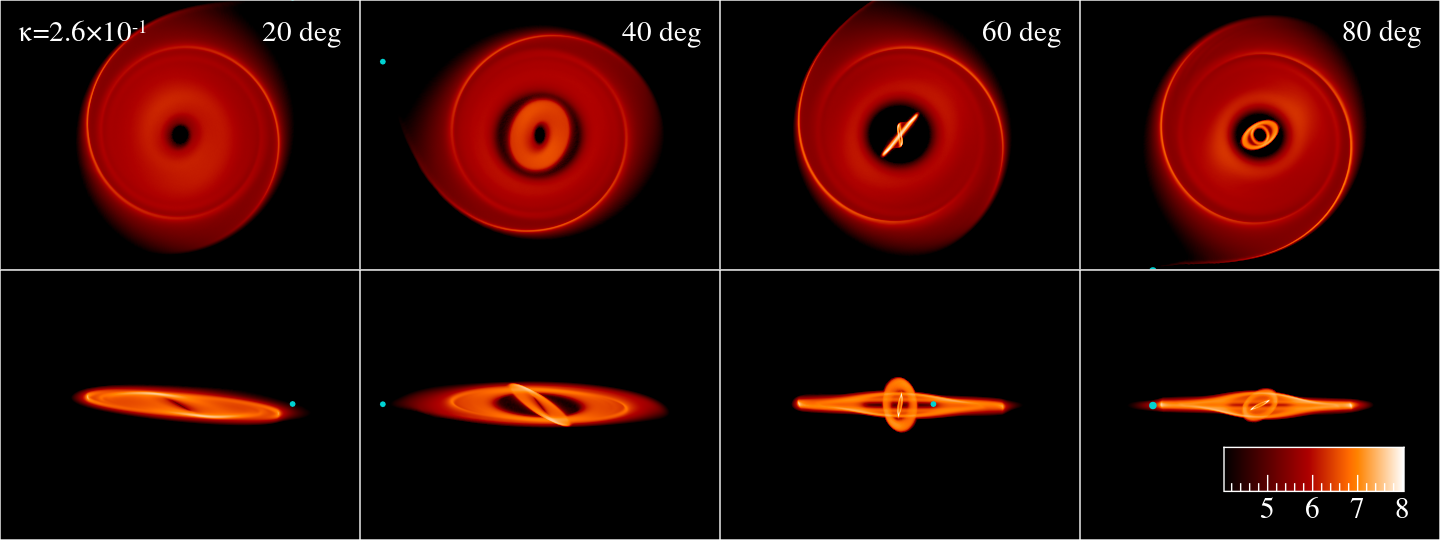}
\caption{Disc breaking in the presence of a binary BH, adapted from \citet{2022MNRAS.509.5608N}. The colour here represents column density in code units. As the disc inclination is increased, the external torques from the BHs overcome the viscous stresses maintaining the disc and it breaks into independently precessing rings.
\label{fig:bec_breaking_BHBs}}
\end{figure}

The misaligned binary scenario was subsequently explored by \citet{2022MNRAS.509.5608N} using 3D smoothed particle hydrodynamic simulations. They spanned the parameter space investigated by \citet{2020MNRAS.496.3060G} and broadly found excellent agreement with the semi-analytic model. Each of their simulations was classified into four outcomes: i) warping, ii) unsuccessful breaking, iii) a single break and iv) multiple breaking. In the latter two cases, the torque applied from the misaligned binary was able to overcome the viscous stresses maintaining the disc causing it to be torn into discrete discs that precessed independently - see Figure~\ref{fig:bec_breaking_BHBs}. Additionally, disc breaking occurred exclusively for parameters that corresponded to the regime above the ‘critical obliquity’ identified by \citet{2020MNRAS.496.3060G}. At inclinations close to the critical obliquity, \citet{2022MNRAS.509.5608N} found examples of ‘unsuccessful breaking’, where the disc showed clear signatures of disc breaking \citep{2018MNRAS.476.1519D} but was unable to fully separate. In the event that gas assisted migration plays a strong role in the evolution of these binaries, we may thus expect discs to be either warped or broken into multiple discrete, precessing sub-discs. 

When the disc is broken and precesses freely, as observed in the simulations of \citet{2022MNRAS.509.5608N}, the back-reaction onto the primary BH was decreased. As a result BHs hosting broken discs did not align their spin as efficiently as those with continuous, warped discs. Using a combination of semi-analytic approaches \citet{2023MNRAS.519.5031S} investigated the inspiral, coalescence and gravitational wave emission of such supermassive binaries taking the results of \citet{2022MNRAS.509.5608N} into account. Under the assumption that disc breaking prevents spin alignment, their results showed distinct sub-populations of aligned BHs that will be observable by LISA. While such observations will provide conclusive evidence for the Bardeen-Petterson effect, \citet{2023MNRAS.519.5031S} also implied that the presence of disc breaking introduces an ambiguity as strongly misaligned BH mergers could be attributed to either a gas-poor environment or a gas-rich environment where disc breaking has occurred. 

State of the art numerical calculations, such as those discussed above, provide a timely opportunity to make robust observational predictions for future missions such as LISA. The  use of realistic accretion disc and circumbinary disc models, with the inclusion of warps and misalignments are vital to understand their impact on spin evolution and thus further work needs to be done to encompass the full, rich and complex behaviour of these systems.
\begin{figure*}
\includegraphics[width=\textwidth]{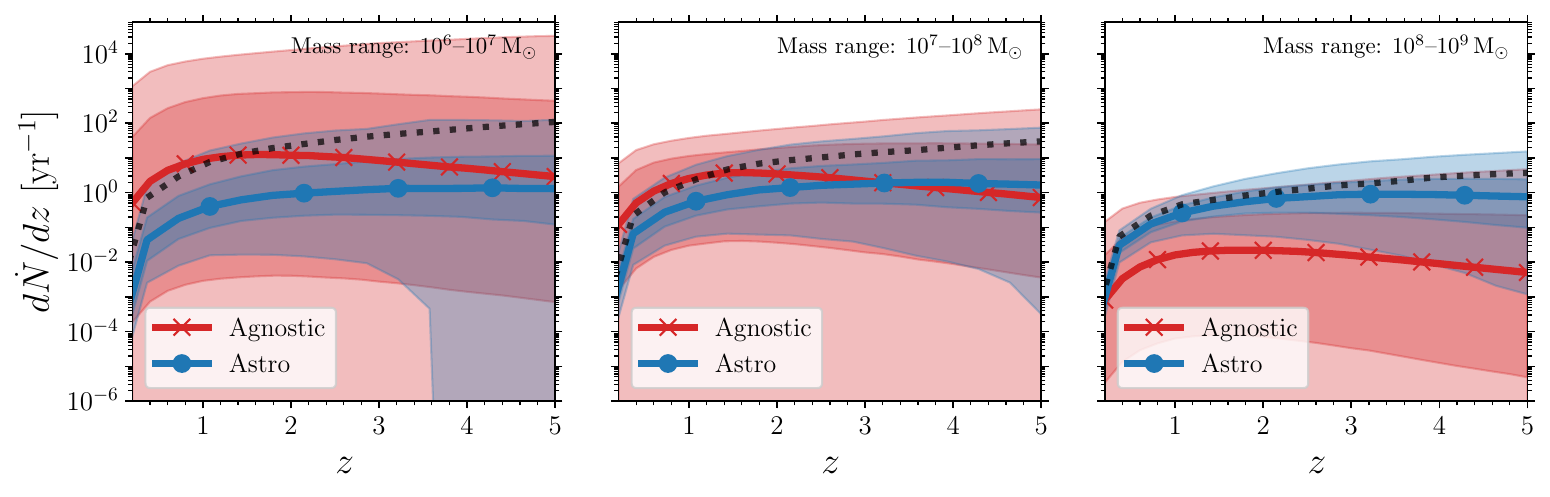}
\caption{\label{fig:GWdNdz}
Assuming the common-noise process observed by the PTAs originates from a population of SMBH binaries, we can make inferences on the properties of the population. 
Here we show posterior distributions for the merger rate per year per unit redshift for three different total mass ranges. 
The red and blue distributions show results for an agnostic and an astrophysically-informed model of the population, respectively. 
The solid lines indicate the median of the posterior, and the dark and light shading indicates the central $50\%$ and $90\%$ credible regions, respectively. 
The dotted line indicates the $99.5\%$ percentile of the astrophysically-informed prior.
Reproduced from~\cite{2023MNRAS.525.2851S}.
}
\end{figure*}

\section{Gravitational waves}
\label{sec:GW}
The currently operating ground-based gravitational wave observatories, such as LIGO~\citep{2015CQGra..32g4001L}, Virgo~\citep{2015CQGra..32b4001A}, and KAGRA~\citep{10.1093/ptep/ptaa125} target frequencies at $\sim 10\,{\rm Hz}$--$1\,{\rm kHz}$.
Ground-based facilities are not able to reach low enough gravitational-wave frequencies to detect supermassive black holes, instead detecting stellar-origin black hole and neutron star \rev{binaries}~\citep{2023PhRvX..13d1039A}.
The future space mission, LISA~\citep[Laser Interferometer Space Antenna][]{2024arXiv240207571C}, will observe the gravitational-wave sky at $0.1\,{\rm mHz}$--$1\,{\rm Hz}$.
Until the arrival of the LISA space mission, the very-low frequency band ($\approx 1\,{\rm nHz}$--$1\,\mu{\rm Hz}$) is 
therefore
the exclusive domain of pulsar timing arrays \citep[PTAs, e.g., ][]{1990ApJ...361..300F,2017arXiv170701615H}. 
While the frequency bands of LISA and PTA do not overlap, SMBH binaries at different stages of their evolution are accessible to both facilities. 

PTA observations consist of long-term radio observations of millisecond pulsars which are used to search for the imprint of gravitational waves in the pulse time of arrivals. 
Focusing on the gravitational-wave background 
\mghk{produced by all the SMBH binaries in the entire Universe}, theoretical predictions for signal amplitude are uncertain~\citep{2013MNRAS.433L...1S,2015MNRAS.451.2417R,2017MNRAS.471.4508K,2020ApJ...897...86C,2022MNRAS.511.5241S}, \mghk{as one will expect after the above discussion.} However, in the last few years, observations from the PTA consortia indicate the presence of a common-noise signal in the datasets of the International PTA~\citep[IPTA,][]{2022MNRAS.510.4873A}, the European PTA~\cite[EPTA,][]{2021MNRAS.508.4970C}, the Parkes PTA~\citep[PPTA,][]{2021ApJ...917L..19G}, and the North American Observatory for Gravitational waves~\citep[NANOGrav,][]{2020ApJ...905L..34A}.
More recently, increased evidence that the observed signal has a gravitational-wave origin has been announced by the EPTA in collaboration with the Indian PTA~\citep[ETPA+InPTA][]{2023A&A...678A..50E, 2022PASA...39...53T}, the PPTA~\citep{2023ApJ...951L...6R}, NANOGrav~\citep{2023ApJ...951L...8A}, the Chinese PTA~\citep[CPTA,][]{2023RAA....23g5024X,CPTA:2016}, and the MeerKAT PTA~\citep[MPTA][]{2023MNRAS.519.3976M,2025MNRAS.536.1489M}.
Assuming a gravitational-wave origin, the significance of the reported signal is between $2\sigma$ and $4\sigma$.
If formed by gravitational waves, the signal could have a number of origins, including SMBH binaries, dark matter, and the early Universe~\citep[e.g.][]{2024A&A...685A..94E, 2023ApJ...952L..37A, 2023ApJ...951L..11A, 2023ApJ...955..132C}.
Further study will be required to confirm the signal~\citep[e.g.][]{2022MNRAS.516..410Z} and the combined datasets of the IPTA will be used.

\rev{In addition to evidence for a gravitational-wave background, the PTA consortia also search for gravitational 
waves from individual SBHB binaries as continuous gravitational waves \citep{Sesana+09,2015MNRAS.451.2417R,Kelley+18,Becsy+22}. 
Candidate signals have been investigated in the latest data sets \citep{EPTA24,Agazie+23}, however further evidence will be required to confirm whether or not these signals originate from individual SBHB binaries. For the purpose of this work, we focus on the evidence for a gravitational-wave background and the population of SMBHs implied by it.}

Throwing caution to the wind, one can make the assumption that the signal observed by the PTAs is indeed the signature of the gravitational-wave background and see what inferences can be placed on an associated population of inspiraling SMBH binaries via a Bayesian analysis. \rev{The results summarised here are presented in full in \citet{2023MNRAS.525.2851S}.}
Following the methods in~\cite{2023MNRAS.525.2851S,2021MNRAS.502L..99M,2019MNRAS.488..401C,2017MNRAS.468..404C,2016MNRAS.455L..72M} which follow the formalism laid out in~\cite{2001astro.ph..8028P}, we use two models of the population; an `agnostic' model which makes minimal assumptions and a `astrophysically-informed' model which uses information from simulation and observation to inform the priors. 
We note that the analysis in~\cite{2023MNRAS.525.2851S} was carried out using the IPTA results in~\cite{2022MNRAS.510.4873A}, however they are consistent with the most recent announcements.
Using the astrophysically-informed model, we can infer merger rate densities at the higher end of the astrophysical prior and shorter delay times between the galaxy merger and black hole binary merger~\citep[\rev{Fig.~\ref{fig:GWdNdz},} see][for full details]{2023MNRAS.525.2851S,2021MNRAS.502L..99M,2019MNRAS.488..401C}.

Turning to future space-based observatories, 
the LISA mission~\citep{2017arXiv170200786A,2024arXiv240207571C} is expected to observe the merger itself for individual SMBH binaries in the gravitational-wave frequency band $0.1\,{\rm mHz}$--$1\,{\rm Hz}$, corresponding to short lived signals from mergers of lower mass SMBH binary systems~\citep[see for example][]{2023LRR....26....2A}.
Parameter estimation studies of merging black hole binaries indicated that LISA inferences on the binary properties might be possible to within $<1\%$ in component mass, dependent on the system.
For futher information see, for example~\cite{2023MNRAS.525.2851S, 2023arXiv230713026P, 2023PhRvD.107l3026P}, which uses the \textsc{Balrog} software package under development for LISA data analysis~\citep{2023MNRAS.522.5358F,2022arXiv220403423K,2021PhRvD.104d4065B,2020ApJ...894L..15R}.

Finally, we consider how PTA and LISA observations will be complementary. 
Again making the assumption that the PTA signal originates from a gravitational-wave background from inspiraling SMBH binaries, we can extrapolate the inference described above to calculate the number of mergers per year of LISA operation as described in~\cite{2023MNRAS.525.2851S}. 
\rev{If the PTA signal does indeed have a SMBH binary origin, the implied population of inspiraling binaries should also provide merging sources in the LISA band.}
The posterior distribution for the merger rate per year per unit redshift is shown in Figure~\ref{fig:GWdNdz} for the agnostic and astrophysically-informed models in red and blue, respectively. 
Again using \textsc{Balrog}, we assess the detectability of these systems to provide an anticipated detection rate for LISA. 
Using the astrophysically-informed model, we predict a $95\%$ upper limit on the detection rate of $\mathcal{R} < 134\,{\rm yr}^{-1}$ for binary masses in the range $10^7$--$10^8\,{\rm M_{\odot}}$. 
For full details of this work see~\cite{2023MNRAS.525.2851S}.
Another study using a semi-analytic galaxy formation model makes similar predictions~\citep{2023arXiv230712245B}.

Further study is required. However, the next few years may well confirm that the PTA observation has a gravitational-wave origin, potentially from a population of SMBH binaries. Looking further ahead, space-based observatories will enable multi-band exploration of these systems through their gravitational waves.


\section{Conclusions}\label{sec:conc}
We have reviewed the state of theoretical modelling and a variety of evidence for the different 
observable effects produced by binary SMBH, ranging from predictions for the cosmological evolution of SMBH, the physics of migration to the common centre after a galaxy merger, \rev{over}
binary SMBH accretion discs to electromagnetic and gravitational wave observations. 
\rev{While direct gravitational wave detections of supermassive black hole mergers are not yet available, the observation of the effects of SMBHs on their immediate environments as well the energetic requirements of active galactic nuclei provide strong arguments for the existence of SMBH in some, if not all centres of galaxies, most likely our own. Cosmological simulations then tell the story of SMBH growth via accretion and SMBH merging. However, uncertainties about the SMBH seeds, namely what the astrophysical processes involved actually are and what masses they produce, and limitations in resolution}
make the \rev{detailed} prediction of SMBH evolution still uncertain. 
\rev{The spin is a key parameter here. While coherent gas accretion, which should occur typically for lower mass SMBH, would tend to produce rapidly spinning black holes, SMBH mergers are expected to produce a slowly spinning remnant.
}
Spin observations via X-ray spectroscopy start informing the picture\rev{.
With a sample of a few dozen SMBH a decline at the highest SMBH masses is seen, consistent with the cosmological expectation of mergers being important in this mass range. M}ore, deep observations, especially towards the higher mass end would be highly desirable. 
Spin evolution and migration in the close binary state will be a result of the interaction with the accretion disc or discs, if one regards the circumbinary and individual accretion discs separately. \rev{Both is still uncertain and depending on modelling parameters like the assumed gas cooling processes. 
Angular momentum vectors of gas and SMBHs will in general not be aligned and for a binary SMBH embedded in a significant amount of gas, breaking of a warped disc can reduce the torques that would otherwise align the SMBH spins with the disc angular momentum vector. Large misalignments of SMBH spins at their merger can, however, produce kicks that may remove the merger remnant from the galactic centre.}

An increasing number of dual AGN is detected in X-rays, radio and via \rev{a range of} other methods. \rev{These observations support the idea that binary
SMBH form during galaxy mergers, with X-ray imaging finding two active black holes within single galaxy structures. High-resolution radio imaging has revealed a good binary SMBH candidate with 7~pc projected separation. Similar discoveries and possibly tracking of orbits seem in reach \citep{2022ApJ...931...12W}. With longer timebases, candidates for sub-parsec binary SMBH are also being discovered from sinusoidal radio light curves
\citep{2022ApJ...926L..35O,2025ApJ...985...59K}. High-resolution optical as well as variability methods, while having to overcome significant issues with noise from unrelated astrophysical processes, have found candidates for sub-pc SMBH binaries.}

Jet precession is a powerful tool here and might act as a magnifying glass for unresolved binaries, or ones where a weakly accreting secondary is only observable via the radio jet which may precess no matter which black hole produces it. 
OJ~287 is an observationally well studied example of a close binary SMBH that deserves further detailed modelling of the complex interplay of discs and 
\rev{parsec-scale} jets.
 Large-scale jet morphologies consistent with geodetic precession expected from sub-parsec separation SMBH binaries have been observed in more than 1000 radio sources. 
 \rev{There is scope for combining AGN observations with future LISA gravitational wave observations \citep{Steinlea24}. Not only may radio emission from jets reveal morphological features even 100s of Myr after the source switched off \citep[e.g.][]{Shabala:2024}, thus providing for a fossil record of jet precession before a merger. Even sources where discs are actively re-aligning, with corresponding jet precession, might be detected with LISA \citep{Steinlea24}.}
 
 \rev{One expects a background of low-frequency gravitational waves produced by all the SMBH binaries in the Universe. A corresponding signal has been seen by pulsar timing arrays. The discussion if this signal is indeed due to the aforementioned SMBH binaries has to be informed by the astrophysics outlined above. A further confirmation would be the detection of the mergers of the putative SMBH population with LISA.}

We have highlighted here some of the best pieces of evidence for binary supermassive black holes. Still more could have been mentioned, including the changes in
stellar mass profile in a galaxy due to the inspiral 
\citep[compare discussion in][]{Krause19} and quasi-periodicities in gamma-ray lightcurves \citep{Rieger:2007,Rieger:2019}. 
Progress in instrumentation and data analysis methods suggest that all these methods have an exciting future.


\begin{acknowledgement}
    This review was inspired by a discussion meeting hosted by the Royal Astronomical Society in London on 14 April 2023.~J.M.S.R acknowledges support from the Science and Technology Facilities Council (STFC) under grant ST/V50659X/1~(project reference 2442592)~and thanks Andy Fabian, Jiachen Jiang and Dom Walton for useful discussions.
    M.A.B acknowledges support from a UKRI Stephen Hawking Fellowship (EP/X04257X/1) and the Science and Technology Facilities Council (STFC), grant code ST/W000997/1.
    H.~M. acknowledges the support of the UK Space Agency, Grant No. ST/V002813/1 and ST/X002071/1. RN acknowledges funding from UKRI/EPSRC through a Stephen Hawking Fellowship (EP/T017287/1). M.G.H.K acknowledges the hospitality of the Institute of Astronomy, University of Cambridge, UK for part of the writing of this manuscript.
\end{acknowledgement}


\bibliography{sample631,smbbh}{}

\end{document}